\documentclass[amssymb,amsmath,superscriptaddress,footinbib]{revtex4}
\usepackage{natbib,graphicx,subfigure,subeqnarray,fancyhdr,amsmath,multirow,color, mathrsfs}
\begin{document}

\renewcommand{\labelitemi}{$-$}
\newcommand{\change}[1]{{\color{black}#1}}
\newcommand{\Fc}{\mathcal{F}}\newcommand{\Rc}{\mathcal{R}}\newcommand{\dd}{\mathrm{d}}
\newcommand{\ee}{\mathrm{e}}\newcommand{\ci}{\mathrm{i}}\newcommand{\ib}{\mathbf{i}}
\newcommand{\jb}{\mathbf{j}}\newcommand{\kb}{\mathbf{k}}\newcommand{\ab}{\mathbf{a}}
\newcommand{\Fb}{\mathbf{F}}\newcommand{\fb}{\mathbf{f}}\newcommand{\Gb}{\mathbf{G}}
\newcommand{\Mb}{\mathbf{M}Ä}\newcommand{\nb}{\mathbf{n}}\newcommand{\Sb}{\mathbf{S}}
\newcommand{\Sbs}{\mathbf{S^*}}\newcommand{\Rb}{\mathbf{R}}\newcommand{\Sigb}{\boldsymbol{\Sigma}}
\newcommand{\Sigbs}{\boldsymbol{\Sigma^*}}\newcommand{\alphab}{\boldsymbol\alpha}
\newcommand{\omegab}{\boldsymbol{\omega}}
\newcommand{\epsb}{\boldsymbol{\epsilon}}
\newcommand{\ub}{\mathbf{u}}
\newcommand{\eb}{\mathbf{e}}\newcommand{\vv}[1]{\underline{#1}}\newcommand{\ev}{\vv{e}}
\newcommand{\rv}{\vv{r}}\newcommand{\TT}[1]{\underline{\underline{#1}}}\newcommand{\omb}{\mathbf{\omega}}
\newcommand{\Ub}{\mathbf{U}}\newcommand{\xb}{\mathbf{x}}\newcommand{\rb}{\mathbf{r}}
\newcommand{\Omegab}{\boldsymbol\Omega}
\newcommand{\ssb}{\mathbf{s}}\newcommand{\Xb}{\mathbf{X}}\newcommand{\Pe}{\mbox{Pe}}\newcommand{\Da}{\mbox{Da}\,}
\newcommand{\mean}[1]{\langle #1\rangle}
\newcommand{\ddp}{[p]^\pm}\newcommand{\taub}{\mbox{\boldmath$\tau$}}\newcommand{\Fr}{\mbox{\textit{Fr}}}
\let\grad\nabla\newcommand{\z}{\zeta}\newcommand{\kk}{\kappa}\newcommand{\tkk}{\tilde{\kappa}}
\newcommand{\e}{\varepsilon}\newcommand{\zb}{\bar{\zeta}}\let\grad\nabla\let\bcdot\cdot
\newcommand{\half}{{\textstyle\frac{1}{2}}}
\newcommand{\textfrac}[2]{{\textstyle\frac{#1}{#2}}}
\newcommand{\LF}[1]{{#1}^{\mathrm{LF}}}\newcommand{\Lap}[1]{{#1}^{\mathrm{L}}}
\newcommand{\ds}{*\!*}\newcommand{\cond}[2]{\frac{\mathrm{D} #1}{\mathrm{D} #2}}
\newcommand{\pard}[2]{\frac{\partial #1}{\partial #2}}\newcommand{\totd}[2]{\frac{\mathrm{d}#1}{\mathrm{d}#2}}
\newcommand{\pardd}[3]{\frac{\partial^2 #1}{\partial #2 \partial #3}}
\newcommand{\Rey}{\mbox{Re}}\newcommand{\Imag}{\mbox{Im}}
\newcommand{\Fpint}{=\!\!\!\!\!\!\!\int}
\newcommand{\txi}{\tilde\xi}\newcommand{\dxi}{\delta\xi}
\newcommand{\tpsi}{\tilde\psi}\newcommand{\dpsi}{\delta\psi}
\makeatletter
\def\sgn{\mathop{\operator@font sgn}}
\makeatother
\def\v{{\vspace{4mm}}}
\title{Phoretic self-propulsion at finite P\'eclet numbers}
\author{S\'ebastien Michelin}
\email{sebastien.michelin@ladhyx.polytechnique.fr}
\affiliation{LadHyX -- D\'epartement de M\'ecanique, Ecole Polytechnique -- CNRS, 91128 Palaiseau, France}
\author{Eric Lauga}
\email{e.lauga@damtp.cam.ac.uk}
\affiliation{Department of Applied Mathematics and Theoretical Physics, University of Cambridge, Cambridge CB3 0WA, United Kingdom}

\date{\today}

\begin{abstract}

Phoretic self-propulsion is a unique example of force- and  torque-free motion on small scales. The classical framework describing the flow field around a particle swimming by self-diffusiophoresis neglects the advection of the solute field by the  flow and assumes that the chemical interaction layer is thin compared to the particle size. In this paper we quantify and characterize the effect of solute advection on the phoretic swimming of a sphere.  We first rigorously derive the regime of validity of the thin-interaction layer assumption at finite  values of the P\'eclet number ($\Pe$). Within this assumption, we solve computationally the flow around Janus phoretic particles and examine the impact of solute advection on propulsion and the flow created by the particle. We demonstrate that although  advection always leads to a decrease of the swimming speed and flow stresslet at high values of the P\'eclet number, an increase can be obtained at intermediate values of $\Pe$. This possible enhancement of swimming depends critically on the nature of the chemical interactions between the solute and the surface. We then derive an  asymptotic analysis of the problem at small  $\Pe$ allowing to rationalize our computational results. Our computational and theoretical analysis is accompanied  by a parallel study of the role of reactive effects at the surface of the particle on swimming (Damk\"ohler number).

\end{abstract}

\maketitle
\section{Introduction}

Self-propulsion at low Reynolds number is usually associated with the biological world. Indeed, many cellular organisms display some form of motility in fluids, in both prokaryotic and eukaryotic worlds \citep{Bray2000}. Since the equations of motion are linear in this regime, the type of motion leading to self-propulsion have to be  non-time-reversible \citep{Purcell1977}. This is typically achieved through the actuation of cellular appendages called flagella or cilia which act on a surrounding viscous fluid in a wave-like fashion \citep{brennen1977,lauga2009}. 

Beyond the biological world, artificial micro-scale swimmers have received increasing attention in recent years, motivated in part by their potential future use in a biomedical context \citep{Nelson2010}. The design of most synthetic swimmers to date has been inspired by that seen in the biological world, and have thus attempted to reproduce for example  the two-dimensional beating of a sperm flagellum \citep{dreyfus2005} or the three-dimensional rotation of a bacterial flagellum \citep{ghosh2009,Zhang2010b,Gao2010}. In all these cases, the actuation is not embedded in the swimmer itself or the surrounding fluid, but is due to the use of external  fields, typically magnetic.

A promising alternative to design truly self-propelled swimmers  takes advantage of the  short-range interaction between the surface of a colloidal particle and an outer field gradient (e.g.~solute concentration, temperature or electric field) to locally create fluid motion in the vicinity of a particle  boundary \citep{anderson1989}. These so-called phoretic mechanisms are responsible for the migration of isotropic  particles  in externally-imposed gradients. Furthermore, they  may  be exploited to generate  self-propulsion when the particle is itself able to generate local gradients, for example through chemical reaction or heat radiation, an idea which has lead to significant activity in the physics and chemistry communities \citep{paxton2004,golestanian2005,golestanian2007,Howse2007,cordova2008,julicher2009b,jiang2010,Ebbens2011}. Such swimmers are usually referred to as phoretic. 

Physically, when the  interaction layer (chemical, electrical, temperature...) is  thin compared to the particle size, phoretic effects amount to generating a distribution of  slip velocities at the particle surface \citep{julicher2009}, and can thus be thought of as a biomimetic analog to the propulsion by dense arrays of short beating cilia \citep{blake1971}.  In order to induce non-trivial tangential chemical gradients and slip velocities, anisotropic  properties of the particle surface are essential for small particles   \citep{golestanian2005,golestanian2007} and isotropic particles cannot swim unless they are large enough for a symmetry-breaking instability involving advection of the surrounding chemical field to take place \citep{michelin2013c}. The main advantage of this type of swimmer design  is the fact that they can swim in the absence of any external field and thus represent true force-free, torque-free self-propulsion   \citep{Wang2009}. The diffusiophoretic propulsion of such solid particles share many similarities with self-propelled Marangoni droplets, which swim as a result of self-generated gradients of reactive surfactants \citep{thutupalli2011,yoshinaga2012,schmitt2013}

In this paper, we  focus  on the  case of self-diffusiophoresis where the slip velocities are induced through the chemical interactions between a diffusive solute and a particle whose surface partly acts as a catalyst for a chemical reaction. A classical continuum framework has been recently proposed to study the dynamics of an isolated phoretic particle through the coupling of a Stokes flow problem to the diffusing and reacting dynamics of the solute \citep{golestanian2007}. 
This original framework was based on three main assumptions: (i)  the diffuse layer where the solute-particle interaction takes place was assumed to be infinitely thin, so that the phoretic effect can be accounted for by a slip velocity at the surface of the particle; (ii) the advection of the solute was neglected,  effectively decoupling the solute diffusion dynamics from the Stokes flow problem; (iii) on the catalyst portion of the particle surface,  the  chemistry was described by a fixed-rate absorption or release of the solute. In recent years, this framework has been used and extended to study a variety of properties of the self-propulsion of asymmetric colloidal particles including advective effects \citep{cordova2008,julicher2009,cordova2013}, the role of geometry \citep{popescu2010}, the impact of more complex surface chemistry \citep{ebbens2012}, or of the non-zero thickness of the interaction layer \citep{sabass2012,sharifimood2013}.

In the coupled fluid-chemical transport problem, neglecting the advection of the solute significantly simplifies the mathematical analysis   since it effectively decouples the two problems. The solute concentration satisfies a diffusion equation which can be solved first, and its solution can then be exploited in the fluid problem to compute the swimming speed and the flow field. This assumption of zero solute P\'eclet number, $\Pe$,  is particularly adapted when the size of the particle is small enough, when 
the particle activity and/or mobility are weak (in some well-defined sense, which will be detailed below) \change{or when the solute diffusivity is large}.  However, advective effects may become significant when the particle is not small compared to  $a_c=D/\mathcal{U}$ with $D$ the solute diffusivity and $\mathcal{U}$ the characteristic phoretic velocity \citep{julicher2009}. \change{This is particularly relevant in the case of large proteins/molecules when $D$ is very small.   The catalytic autodegradation of hydrogen peroxyde by platinum Janus particles, and the resulting locomotion through gradients of oxygen, corresponds to a critical size $a_c\approx 10$--$100\mu$m \citep{Howse2007,ebbens2012}}. Large $\Pe$ values can also be obtained at much smaller scales when the chemical species correspond to larger molecules \citep[such as surfactants,][]{thutupalli2011}.  When $\Pe$ is no longer negligible, advection of the solute has been shown to significantly impact the velocity of such particles \citep{julicher2009,michelin2013c,khair2013}. Furthermore, the validity of the slip-velocity assumption in the presence of strong advective effects needs to be  investigated. Solute advection at large values of $\Pe$ will  lead to chemical boundary layers and not only can advection within the diffuse layer   become significant but  the diffuse layer thickness might no longer be negligible  compared to the concentration boundary layer.

In addition, most self-diffusiophoresis studies consider either a fixed absorption release of solute at their surface \citep{golestanian2007,julicher2009,sabass2012} or a one-step absorption reaction where the solute flux is proportional to its local concentration \citep{cordova2008}. The former approach can be seen as a particular case of the latter where the solute concentration is only weakly impacted by the reaction and remains mainly set by its far-field value, a  limit  particularly adapted to that of  small particles. The importance of reactive effects, measured by the Damk\"ohler number $\Da$, must still be quantified. For large values of $\Da$, solute diffusion is too slow to refresh the solute content of the fluid near the  surface of the particle and phoretic effects may be reduced, potentially  impacting self-propulsion.

The goal of the present paper is to  quantify and characterize the effect of advection and reaction on  phoretic self-propulsion. We first introduce the general continuum phoretic model for a spherical particle with arbitrary surface chemical properties. Asymptotic expansions are then exploited  to  analyze in detail the validity of the thin-interaction layer limit  in the presence of advective and reactive effects. Using this assumption, a mathematical and computational framework is obtained for solving the phoretic problem at arbitrary values of $\Pe$ and $\Da$ for axisymmetric particles, in particular for  Janus particles possessing one   chemically-active cap while the rest of the particle surface is chemically inert \citep{walther2008}. We then use computations to  analyze the impact of advection and reaction on the swimming velocity and the flow field induced by the  particle motion. We show in particular that advective effects can increase the swimming speed of phoretic swimmers and amplify the flow they induce in the far field. We finally use analytical calculations at small $\Pe$ and $\Da$ numbers to rigorously  calculate   the sensitivity to both advection and reaction of arbitrary Janus particles, thereby explaining our computational results.

\section{Autophoretic propulsion}
\label{sec:general}
The dynamics of an isolated solid particle of radius $a$ \change{are considered} in a fluid of density $\rho_f$ and dynamic viscosity $\eta_f$. A solute $S$ dispersed in the fluid and characterized by its  concentration, $C(\xb,t)$,  interacts with the particle's surface through a short range potential $\Phi(\xb)=k_BT\phi(\xb)$ where $\lambda$ is the range of the interaction potential (i.e.~$|\Phi|/k_BT\ll 1$ if $|\xb|-a\gg \lambda$, where the centre of the coordinate system is taken to be at the centre of the sphere). The solute is characterized by a far-field concentration $C_\infty$ and may be released and/or absorbed at the surface through chemical reaction. 

In the following, a general framework \change{is presented} that can account for two different types of surface chemistry \citep{michelin2013c}:  a fixed-flux absorption/release characterized by an activity $\mathscr{A}$  or  a fixed-rate one-step chemical reaction $S\rightarrow P$ characterized by a reaction rate $\mathscr{K}$. In the latter case, although both reactant $S$ and product $P$ may interact with the surface, we will neglect for simplicity the interaction with $P$ but what follows may easily be generalized to account for the chemical interaction of both species with the surface.  The chemical properties of the particle surface are then characterized either by a distribution of  activity or of reaction rate. 

The solute $S$ is assumed to  diffuse with diffusivity $D$ and to be advected by the fluid flow. In the following, \change{it is assumed} that the Reynolds number $\Rey=\rho_fUa/\eta_f$ is small enough for both fluid and solid inertia to be negligible. If furthermore the  particle density \change{is taken}  equal to that of the fluid, then the particle is force-free and torque-free. 

Near the particle, the interaction between the solute in suspension and the particle surface induces \change{a force $-\grad\Phi$ on a given solute molecule. As a result, the force density applied on the fluid is} $-C(\xb)\grad\Phi$. In a reference frame attached to the centre of the  particle, the equations of motion for the fluid flow simplify to Stokes' equations
\begin{equation}\label{eq:stokes}
{\bf 0}=-\grad p + \eta_f\nabla^2\ub-C\grad\Phi,\quad \grad\cdot\ub=0,
\end{equation}
subject to the far-filed condition and the no-slip boundary condition on the particle surface ($r=a$ with $r=|\xb|$) 
\begin{equation}\label{BC}
\ub(r\rightarrow\infty)\sim-\Ub-\Omegab\times\xb, \quad \ub(r=a)=0.
\end{equation}
In Eq.~\eqref{BC}, $\{\Ub,\Omegab\}$ are the unknown translation and rotation velocity of the rigid particle, respectively. 

The advection-diffusion for the solute is governed by the following equation
\begin{equation}
\pard{C}{t}+\grad\cdot\mathbf{j}=0,
\end{equation}
where the solute flux $\mathbf{j}$ includes advection by the flow,  diffusion,  and transport by the interaction potential as
\begin{equation}
\mathbf{j}=C\ub-D\left(\grad C+\frac{C\grad \Phi}{k_BT}\right)\cdot
\end{equation}
The solute concentration must also satisfy the far-field condition,
\begin{equation}
C(r\rightarrow\infty)\sim C_\infty.
\end{equation}
Finally,  the chemical properties of the particle surface control the surface flux. Denoting  $\xb^S$ a point on the surface of the particle, we have 
\begin{equation}
\left.D\nb\cdot\left(\grad C+\frac{C\grad\Phi}{k_BT}\right)\right|_{(|\xb^S|=a)}=\left\{\begin{array}{l}-\mathscr{A}(\xb^S)\quad \textrm{(fixed-flux);}\\ \\\mathscr{K}(\xb^S)C\quad \textrm{(fixed-rate).} \end{array}\right. 
\end{equation}
In the fixed-flux case, a positive activity corresponds to an emission of solute while negative activity corresponds to absorption. The one-step chemical reaction with fixed rate always corresponds to an absorption and can hence be seen as a negative activity depending on the local concentration.  Note that we assume the  basic mechanism of absorption/desorption of the reactant/product on the surface catalyst  to be fast enough that the solute concentration on the surface is at equilibrium with its immediate fluid environment at all times. 

This set of equations for $\{C,\ub,p\}$ \change{and $(\Ub,\Omegab)$} is closed by imposing \change{the force- and torque-free conditions on the particle}
\begin{align}\label{sumF}
\int\!\!\!\int_{r=a}\boldsymbol\sigma\cdot\nb\,\dd S&+\int\!\!\!\int\!\!\!\int_{\Omega_f}C\grad\Phi \dd\Omega=0,\\
\int\!\!\!\int_{r=a}\xb\times(\boldsymbol\sigma\cdot\nb)\,\dd S&+\int\!\!\!\int\!\!\!\int_{\Omega_f}\xb\times (C\grad\Phi) \dd\Omega=0.\label{sumT}
\end{align}
In Eq.~\eqref{sumF} the total force applied on the particle is the sum of the hydrodynamic force, with stress tensor $\boldsymbol\sigma=-p\mathbf{1}+\eta_f(\grad\ub+\grad\ub^T)$,  and of the interaction forces with the solute in the entire fluid domain. Since the solute-surface interaction is short-ranged, this is approximately  equivalent to imposing zero-hydrodynamic force and torque conditions on any surface outside the interaction layer, e.g.~a sphere of radius $R \gtrsim \lambda + a$.

Both the fixed-rate and fixed-flux approaches can be combined into a single framework by solving for $c=C-C_\infty$ instead of $C$. The advection diffusion problem is now written as
\begin{align}
\pard{c}{t}+\ub\cdot\grad c&=D\grad\cdot\left(\grad c+\frac{(c+C_\infty)\grad\Phi}{k_BT}\right),\\
c(r\rightarrow\infty)&\rightarrow 0,\\
D\nb\cdot\left(\grad c+\frac{(c+C_\infty)\grad\Phi}{k_BT}\right)&=-\mathscr{A}^*+\mathscr{K}c\qquad \textrm{for   }|\xb^S|=a.
\end{align}
The fixed-flux approach is obtained for $\mathscr{K}=0$ and $\mathscr{A}^*=\mathscr{A}$, while in the fixed-rate approach $\mathscr{K}\neq 0$ and $\mathscr{A}^*=-\mathscr{K}C_\infty$.

The problem is non-dimensionalized using $a$ as characteristic length. Since $c$ is the concentration distribution relative to its far-field value,  its characteristic variations scale with the normal gradients imposed at the  surface by  chemistry. A natural scale for $c$ is therefore $[c]=\mathcal{A}a/D$ with $\mathcal{A}$ the typical magnitude of the modified activity (either given by the magnitude of $\mathscr{A}(\xb^S)$ or by $\mathcal{K}C_\infty$ with $\mathcal{K}$ the magnitude of $\mathscr{K}$). A characteristic scale for the velocity $\ub$ is obtained from the dominant balance in the diffuse layer between viscous diffusion and solute-surface interactions and is chosen as $[U]=k_BT\lambda^2[c]/\eta_f a$, from which the characteristic pressure is obtained as $k_BT[c](\lambda/a)^2$ and characteristic time as $a/[U]$. The phoretic propulsion problem above now becomes in dimensionless form
\begin{align}
\nabla^2\ub-\grad p&=\frac{(c+c_\infty)\grad\phi}{\varepsilon^2},\quad \grad\cdot \ub=0,\label{eq:stokesnd}\\
\Pe\left(\pard{c}{t}+\ub\cdot\grad c\right)&=\grad\cdot\left(\grad c+(c+c_\infty)\grad\phi\right),\label{eq:advdiffnd}\\
c(r\rightarrow\infty)\rightarrow 0,&\qquad \ub(r\rightarrow\infty)\sim-(\Ub+\Omegab\times\xb),\label{eq:bcnd1}\\
\ub\Big|_{|\xb^S|=1}=\mathbf{0},&\qquad \nb\cdot\Big[\grad c+(c+c_\infty)\grad\phi\Big]_{|\xb^S|=1}=k(\xb^S)+\Da k(\xb^S)c.\label{eq:bcnd2}
\end{align}
and is characterized by four non-dimensional numbers:
\begin{equation}
\Pe=\frac{k_BT\mathcal{A}\lambda^2 a}{\eta_f D^2},\quad \Da=\frac{\mathcal{K}a}{D},\quad \varepsilon=\frac{\lambda}{a},\quad c_\infty=\frac{DC_\infty}{\mathcal{A}a}\cdot
\end{equation}
\change{Note that in \eqref{eq:bcnd2}, $k(\mathbf{x}^S)$ is defined as $-\mathscr{A}^*/\mathcal{A}$ for both the fixed-flux and fixed-rate problems.}
The P\'eclet number, $\Pe$, is the ratio of diffusive to advective timescales, the Damk\"ohler number, $\Da$, is the ratio of  diffusive to reactive timescales, $\varepsilon$ is the dimensionless range of the interaction potential, and $c_\infty$ is the ratio of the far-field concentration to the typical variations of concentrations around the particle. Note that the equations above are valid even when $\varepsilon$ is not small. In the next section, we consider the classical thin-layer limit, $\varepsilon \ll 1$, for finite values of both $\Pe$ and $\Da$.

\section{The thin-interaction layer limit and its limitations}
\label{sec:thinlayer}

Most studies on self-diffusiophoresis focus on the $\varepsilon\ll1$ limit of short-ranged potential, when the solute-particle interactions are restricted to a thin boundary layer around the particle  \citep{golestanian2005,golestanian2007}.  In this thin-interaction layer limit, all phoretic effects are bundled into two boundary conditions applied on the outer boundary of the interaction layer which is identical to  the particle surface in the limit $\varepsilon\ll 1$, namely    a slip velocity due to tangential solute gradients and a normal solute flux imposed by the chemistry at the particle surface.  In this section, we revisit this limit of short-range potential $\varepsilon\ll 1$, in order to investigate the validity of that framework when neither advection ($\Pe$) nor reaction ($\Da$) can be neglected. \change{Diffusiophoresis shares several fundamental properties and mechanisms with other phoretic phenomena \citep{anderson1989}, and it should be noted that the ``thin interaction layer'' analysis discussed below shares many similarities with the ``thin Debye layer'' limit considered in  classical work on electrophoresis \citep[]{obrien1983,prieve1984}. Recently, Ref.~\citep{yariv2010} proposed a detailed analysis of the asymptotic regime in the case of electrophoresis of particles in externally-imposed electric fields.}

The main result of this section is to show that the validity conditions for each of the boundary conditions above correspond to two distinct mathematical limits. First, in the limit $\varepsilon^2 \Pe\ll 1$, the flow   outside the interaction layer can be solved for taking into account a slip velocity $\ub^S$ at the boundary given by
\begin{equation}
\ub^S=M(\mathbf{I}-\nb\nb)\cdot\grad c,\label{eq:slip}
\end{equation}
with the local mobility $M$ defined from the local interaction potential profile. If additionally we have $\varepsilon\Pe\ll 1$  then advection within the interaction layer is negligible, and the solute advection-diffusion outside this interaction layer can be solved for independently from the interaction layer dynamics by applying on the outer boundary of this layer the flux condition imposed by the chemistry at the particle surface. 

Since we have $\varepsilon\ll1$, then in order for both results to be valid we need to be in the limit $\varepsilon\Pe\ll 1$. In the rest of Section~\ref{sec:thinlayer} we  present  the technical derivation of these two  conditions and readers mostly interested in the particle  dynamics may easily skip these derivations, retaining only the two conclusions above. Note that the derivations and results presented in this section are valid regardless of the surface properties of the spherical particle (activity and interaction potential). In particular, they are applicable to both axisymmetric and non-axisymmetric distributions. In the following, $\zeta=(\theta,\phi)$ generically stands for the two angular coordinates in spherical polar coordinates and is used to characterize this angular (and not necessarily axisymmetric) dependence of the particle's properties.

The derivations below follow the classical approach of matched asymptotic expansions \citep{bender1978}, distinguishing between an outer solution, for which $r-1=O(1)$ and corresponding to the region where solute-particle interactions are negligible, and an inner solution, obtained for $\rho=(r-1)/\varepsilon=O(1)$ in the interaction layer \citep{brady2011,sabass2012,sharifimood2013}.

\subsection{Outer solution}
Expanding all outer fields in the form of regular expansion in $\varepsilon$, $f=f_0(r,\zeta)+\varepsilon f_1(r,\zeta)...$,  and provided that $\phi_0=\phi_1=0$, which is expected for all classical interaction potential decaying at least as fast as $1/(r-1)^2$, the outer problem becomes at leading order
\begin{align}
\nabla^2\ub_0-\grad p_0&=0,\quad \grad\cdot \ub_0=0,\label{eq:stokes0}\\
\Pe\left(\pard{c_0}{t}+\ub_0\cdot\grad c_0\right)&=\nabla^2c_0,\\
c_0(r\rightarrow\infty)\rightarrow 0,&\qquad \ub_0(r\rightarrow\infty)\sim-\Ub_0-\Omegab_0\times\xb,\label{eq:farfield0}
\end{align}
and is identical to other advection-diffusion problems in Stokes flow, such as the feeding  of model ciliates \citep{michelin2011,michelin2013b}. First-order corrections in $\varepsilon$ (namely all  $f_1$ quantities) satisfy the exact same equations. For both problems, the boundary conditions at $r=1$ must be obtained through matching with the inner solution by expanding the different fields for $r-1\ll 1$ as
\begin{equation}
f(r,\zeta)=f_0(1,\zeta)+(r-1)\pard{f_0}{r}(1,\zeta)+\varepsilon f_1(1,\zeta)+o((r-1),\varepsilon).\label{eq:outerexp}
\end{equation}

\subsection{Inner solution}
We now focus on the inner problem for $\rho=(r-1)/\varepsilon=O(1)$. Defining $\tilde\phi(\rho,\zeta)=\phi((r-1)/\varepsilon,\zeta)$, we can write  $\tilde\phi$ as a regular expansion in $\varepsilon$ as
\begin{equation}
\tilde\phi(\rho,\zeta)=\tilde\phi_0(\rho,\zeta)+\varepsilon\tilde\phi_1(\rho,\zeta)...,
\end{equation}
and the same expansion can be carried out for the inner concentration $\tilde c$, and all velocity components. Anticipating on the dominant balance in the momentum equations, the inner pressure is expanded as
\begin{equation}
\tilde p(r,\mu)=\frac{\tilde p_0}{\varepsilon^2}+\frac{\tilde p_1}{\varepsilon}+\tilde p_2+...
\end{equation}
Note that this difference of scaling between the inner and outer pressures impose that $\tilde p_i(r\rightarrow\infty)\rightarrow 0$ for $i=0$ and $1$.

We now substitute  these expansions into Stokes equations, Eq.~\eqref{eq:stokesnd}. Keeping only the first two dominant terms, they can be rewritten as
\begin{align}
&\pard{\tilde u_{r0}}{\rho}+\varepsilon\left[\pard{\tilde u_{r1}}{\rho}+2\tilde u_{r0}+\grad\cdot\tilde\ub_{\parallel 0}\right]=O(\varepsilon^2),\label{eq:continuityinner}\\
\pard{\tilde p_0}{\rho}&+(c_\infty+\tilde c_0)\pard{\tilde\phi_0}{\rho}+\varepsilon\left[\pard{\tilde p_1}{\rho}+(c_\infty+\tilde{c}_0)\pard{\tilde\phi_1}{\rho}+\tilde c_1\pard{\tilde\phi_0}{\rho}-\pard{^2\tilde{u}_{r0}}{\rho^2}\right]=O(\varepsilon^2),\label{eq:stokesrinner}\\
\pard{^2\tilde\ub_{\parallel 0}}{\rho^2}-\grad_\parallel\tilde{p}_0&-(c_\infty+\tilde{c}_0)\grad_\parallel\tilde{\phi}_0+\varepsilon\left[\pard{^2\tilde\ub_{\parallel 1}}{\rho^2}+2\pard{\tilde\ub_{\parallel 0}}{\rho}-\grad_\parallel \tilde{p}_1-\tilde{c}_1\grad_\parallel\tilde{\phi}_0-(c_\infty+\tilde{c}_0)\grad_\parallel\tilde{\phi}_1\right.\nonumber\\
&\left.+\rho\grad_\parallel\tilde{p}_0+\rho(c_\infty+\tilde{c}_0)\grad_\parallel\tilde{\phi}_0\right]=O(\varepsilon^2),\label{eq:stokesthinner}
\end{align}
with   $\ub_\parallel=(\mathbf{I}-\eb_r\eb_r)\cdot\ub$ and $\grad_\parallel p=(\mathbf{I}-\eb_r\eb_r)\cdot\grad p$.
Similarly, the advection-diffusion problem for the inner concentration $\tilde{c}$ can be rewritten as
\begin{align}
\pard{}{\rho}\left(\pard{\tilde c_0}{\rho}+(c_\infty+\tilde c_0)\pard{\phi_0}{\rho}\right)+&\varepsilon\left[\pard{}{\rho}\left(\pard{\tilde c_1}{\rho}+\tilde c_1\pard{\phi_0}{\rho}+(c_\infty+\tilde c_0)\pard{\phi_1}{\rho}\right)+2\left(\pard{\tilde{c}_0}{\rho}+(c_\infty+\tilde c_0)\pard{\tilde\phi_0}{\rho}\right)\right]\nonumber\\
&=\varepsilon\Pe\,\tilde u_{r0}\pard{\tilde c_0}{\rho}+\varepsilon^2\Pe\left(\tilde u_{r1}\pard{\tilde c_0}{\rho}+\tilde{\ub}_{\parallel 0}\cdot\grad_\parallel\tilde{c}_0\right)+O(\varepsilon^2,\varepsilon^3\Pe),\label{eq:advdiffinner}
\end{align}
with boundary condition at the sphere surface, Eq.~\eqref{eq:bcnd2}, becoming
\begin{equation}
\pard{\tilde c_0}{\rho}+(c_\infty+\tilde c_0)\pard{\tilde\phi_0}{\rho}+\varepsilon\left[\pard{\tilde c_1}{\rho}+(c_\infty+\tilde c_0)\pard{\tilde\phi_1}{\rho}+\tilde c_1\pard{\tilde\phi_0}{\rho}\right]=\varepsilon k(\zeta)(1+\Da\tilde{c}_0)+O(\varepsilon^2),\label{eq:advdiffbcinner}
\end{equation}
at $\rho=0$.

At leading order, mass conservation, Eq.~\eqref{eq:continuityinner}, imposes that $\partial\tilde{u}_{r0}/\partial \rho=0$. Together with the impermeability condition at the particle boundary, this shows that 
\begin{equation}
\tilde u_{r0}(\rho,\zeta)=0.
\end{equation}
At leading order, momentum conservation in the radial and azimuthal directions, Eqs.~\eqref{eq:stokesrinner}--\eqref{eq:stokesthinner}, lead to, 
\begin{align}
\pard{\tilde{p}_0}{\rho}&+(c_\infty+\tilde c_0)\pard{\tilde \phi_0}{\rho}=0,\label{eq:ur1}\\
\pard{^2\tilde\ub_{\parallel 0}}{\rho^2}&-\grad_\parallel\tilde{p}_0-(c_\infty+\tilde{c}_0)\grad_\parallel\tilde{\phi}_0=0.\label{eq:uth1}
\end{align}
Provided that $\varepsilon^2\Pe\ll 1$, Eqs.~\eqref{eq:advdiffinner} and \eqref{eq:advdiffbcinner} can be solved to leading order as
\begin{equation}
\tilde c_0(\rho,\zeta)=-c_\infty+\mathcal C_0(\zeta)\ee^{-\tilde\phi_0(\rho,\zeta)}.\label{eq:c0in}
\end{equation}
Substitution of this result into Eq.~\eqref{eq:ur1} together with the decay condition of $\tilde p_0$ and $\tilde\phi_0$ for $\rho\gg 1$ leads to
\begin{equation}
\tilde p_0(\rho,\zeta)=\mathcal{C}_0(\zeta)(\ee^{-\tilde\phi_0(\rho,\zeta)}-1).
\end{equation}
Finally, substituting this result into Eq.~\eqref{eq:uth1}, we obtain after integration and rearrangement,
\begin{equation}
\tilde \ub_{\parallel 0}(\rho,\zeta)=-\grad_\parallel\mathcal{C}_0\left[\int_0^\infty R\left(\ee^{-\tilde\phi_0(R,\zeta)}-1\right)\dd R+\int_\rho^\infty(\rho-R)\left(\ee^{-\tilde\phi_0(R,\zeta)}-1\right)\dd R\right]+\boldsymbol\beta\rho,
\end{equation}
where $\mathcal{C}_0(\zeta)$ and $\boldsymbol\beta(\zeta)$ are to be determined through matching with the outer solution.

\subsection{Matching at leading order - Slip velocity}
Matching the outer and inner solutions at leading order shows that $\boldsymbol\beta=0$ and provides the following relations
\begin{align}
&c_0(1,\zeta)=\mathcal{C}_0(\zeta)-c_\infty,\\
&u_{r0}(1,\zeta)=0,\label{eq:match2}\\
&\ub_{\parallel 0}=M(\zeta)\grad_\parallel\mathcal{C}_0.\label{eq:match3}
\end{align}
with $M$, the mobility coefficient, given by
\begin{equation}
M(\zeta)=-\int_0^\infty\rho\left(\ee^{-\tilde\phi_0(\rho,\zeta)}-1\right)\dd \rho.\label{eq:mobility}
\end{equation}

Combining Eqs.~\eqref{eq:match2}--\eqref{eq:match3}, we therefore establish that, provided $\varepsilon^2\Pe\ll 1$, the outer problem can be solved at leading order using a slip boundary condition 
\begin{equation}
\ub=M(\mathbf{I}-\nb\nb)\cdot\grad c\quad \textrm{at}\quad r=1\label{eq:slipvelocity},
\end{equation}
characterized by the mobility coefficient $M(\zeta)$ in Eq.~\eqref{eq:mobility} \citep{anderson1989}. For locally attractive interactions ($\phi_0<0$), the mobility coefficient is negative and the slip velocity is oriented down-gradient, while for locally repulsive interactions ($\phi_0>0$) the slip velocity is oriented in the direction of the tangential solute gradient.

\subsection{Validity of the flux condition}
Since the flux boundary condition in Eq.~\eqref{eq:bcnd2} does not appear at leading order, the slip velocity result, Eq.~\eqref{eq:slipvelocity}, is  not sufficient to close the outer system formed by Eqs.~\eqref{eq:stokes0}--\eqref{eq:farfield0}.   In order to obtain the additional boundary condition on $c_0$, it is necessary to carry out the expansion in the inner region to the next order \change{\citep[see also the work of][for a similar treatment in the case of electrophoresis]{yariv2010}}. Provided $\varepsilon\Pe\ll 1$, the next order contribution to the advection-diffusion equation and boundary condition, Eqs.~\eqref{eq:advdiffinner}--\eqref{eq:advdiffbcinner}, leads to
\begin{equation}
\pard{\tilde c_1}{\rho}+\tilde c_1\pard{\tilde\phi_0}{\rho}=-(c_\infty+\tilde c_0)\pard{\tilde\phi_1}{\rho}+k(\zeta)+\Da k(\zeta)\tilde{c}_0.
\end{equation}
After integration, we obtain
\begin{align}
\tilde c_1(\rho,\zeta)=\ee^{-\tilde\phi_0(\rho,\zeta)}&\Big[\mathcal{C}_1(\zeta)+\rho k(\zeta)\Big(1+\Da (\mathcal{C}_0(\zeta)-c_\infty)\Big)+k(\zeta)(\Da c_\infty-1)\int_\rho^\infty\left(\ee^{\tilde\phi_0(R,\zeta)}-1\right)\dd R-\mathcal{C}_0(\zeta)\tilde\phi_1\Big]. 
\end{align}
Using the previous equation and Eq.~\eqref{eq:c0in}, in the limit $\rho\gg 1$ and $\varepsilon\ll 1$, we have 
\begin{equation}
\tilde c=\mathcal{C}_0(\zeta)-c_\infty+\varepsilon\rho\left[k(\zeta)+\Da k(\zeta)\left(\mathcal{C}_0-c_\infty\right)\right]+\varepsilon\mathcal{C}_1(\mu)+o(\varepsilon,\varepsilon \rho).
\end{equation}
Matching with the expansion of the outer solution $c$ when $(r-1)\ll 1$, Eq.~\eqref{eq:outerexp}, we  obtain
\begin{equation}
\pard{c_0}{r}(1,\zeta)=k(\zeta)+\Da k(\zeta)c_0(1,\zeta).
\end{equation}
This equation simply states that the diffusive flux at the outer boundary of the interaction layer is equal to the diffusive flux at the particle surface. In this limit,  advection is negligible in the interaction layer and the solute simply diffuses in the radial direction. This provides the missing boundary condition for $(\ub_0,c_0)$ and leads to an autonomous and well-posed set of equations. 

This condition however breaks down when advection within the diffuse layer becomes important and $\varepsilon\Pe=O(1)$, or equivalently when $k_BT\mathcal{A}\lambda^3/\eta_f D^2=O(1)$. That condition does not depend on the size $a$ of the phoretic particle, but only on the surface properties and diffusivity coefficients. Current experimental applications correspond to interaction layers of typical thickness $\lambda\lessapprox 1$nm \citep{Howse2007,ebbens2012}, so that $\varepsilon=O(10^{-5}$--$10^{-3})$ for micrometric particles, and therefore advective effects  {within the interaction layer} are indeed negligible, even for $\Pe=O(1)$.

Note that even in the absence of advective effects in the outer region ($\Pe=0$), advective effects within the interaction layer may modify significantly the diffusive flux when the interaction potential is strong enough for the adsorption length to be comparable to the particle's size \citep{anderson1991}. Such effects are \change{implicitly neglected here: our choice for the scaling of the flow velocity within the interaction layer  assumes that the adsorption length and interaction layer thickness are comparable.}

To conclude, it is  noteworthy that the validity conditions  for the two approximations resulting from the thin diffuse layer framework, namely the slip velocity and the boundary flux of solute, are mathematically different at high $\Pe$, namely $\varepsilon\ll \Pe^{-1/2}$ for the slip velocity definition vs.~$\varepsilon\ll \Pe^{-1}$ for the boundary flux of solute.

\section{Self-propulsion of autophoretic Janus particles at finite P\'eclet and Damk\"ohler numbers}
\label{sec:model}

In  Sections~\ref{sec:model}--\ref{sec:janus_general}, we present a model for the autophoretic self-propulsion of axisymmetric particles based on these approximations and investigate the effect of $\Pe$ on the self-propulsion properties of autophoretic particles.  We assume that $\varepsilon$ is sufficiently  small such that  the limit $\varepsilon\Pe\ll 1$ allows us to consider intermediate and large values of $\Pe$. In that limit, the solute-particle interactions are  entirely accounted for through a slip velocity, Eq.~\eqref{eq:bcnd2}, and  a flux condition, Eq.~\eqref{eq:slip}, both valid at its outer limit ($r=1^+$). 

Focusing on steady-state propulsion, the resulting phoretic problem is expressed as
\begin{align}
&\nabla^2\ub-\grad p=0,\quad \grad\cdot \ub=0,\label{eq:phoretic_01}\\
&\Pe\,\ub\cdot\grad c=\nabla^2c\,\label{eq:phoretic_02},\\
&c\rightarrow 0\quad\textrm{and}\quad \ub\sim-(\Ub+\Omegab\times\xb)\quad\textrm{for}\quad r\rightarrow\infty\label{eq:phoretic_03},\\
&\pard{c}{r}=k(\zeta)(1+\Da c)\quad\textrm{and}\quad\ub=M(\zeta)(\mathbf{I}-\nb\nb)\cdot\grad c\qquad\textrm{for}\quad r=1,\label{eq:phoretic_04}
\end{align}
The  swimming velocity and  rotation rate  are obtained using the reciprocal theorem for a force-free and torque-free particle \citep{stone1996} and given by
\begin{equation}
 \mathbf{U}=-\frac{1}{4\pi}\int_{r=1}\ub_{\parallel}\dd S,\qquad \Omegab=-\frac{3}{8\pi}\int_{r=1}\nb\times\ub_{\parallel}\dd S\label{eq:phoretic_05}.
\end{equation}

The problem now depends only on two dimensionless parameters,  $\Da$ and $\Pe$.  The Damk\"ohler number  characterizes the importance of diffusion in controlling the surface kinetics of the solute. When $\Da=0$, diffusion is fast enough for the absorption of solute to be controlled by its far-field concentration and to be essentially independent on the local fluctuations of solute concentration (fixed-flux framework). In contrast, for finite $\Da$, the concentration fluctuations resulting from the absorption of solute at the interface are significant. The P\'eclet number, $\Pe$, characterizes the relative importance of advection and diffusion on the solute distribution. When $\Pe=0$, the flow resulting from phoretic effects at the particle surface has no impact on the solute distribution. Both non-dimensional numbers can also be seen as a measure for the particle size and the classical framework ($\Pe=\Da=0$) is therefore appropriate for small particles. Here, we thus investigate the advective and reactive effects when the particle size is no longer  small enough for  both $\Pe$ and $\Da$ to be neglected.

For simplicity, we exclusively focus on the absorption of a solute through chemical reaction ($k>0$), and the  ``release'' problem is easily obtained from our results by changing $M$ into $-M$. We also assume in the main text that the mobility is uniform, i.e.~that the interaction potential is isotropic around the sphere, $\phi(\xb)=\phi(|\xb|)$. However, as we show in  Appendix \ref{sec:nonunif_mob}, our framework can also be used in the case of non-uniform mobility, and leads to a generalization of the results presented in the main text to arbitrary mobility distributions. Finally, we note that the magnitude of $M$ effectively determines the characteristic velocity outside the diffuse layer. Therefore, it is more relevant to rescale the velocity (and pressure) so as to include the effect of the potential distribution. The characteristic velocity scale is now
\begin{equation}
[U]=\frac{k_BT\lambda^2[c]}{\eta_f a}\left|\int_0^\infty\rho\left(\ee^{-\tilde\phi_0(\rho)}-1\right)\dd \rho\right|, 
\end{equation}
so that the non-dimensional mobility is simply $M=\pm 1$.

\subsection{The axisymmetric phoretic problem}
From Eq.~\eqref{eq:phoretic_05}, we see that a sufficient condition for the self-propulsion of a phoretic particle relies on its ability to generate a slip velocity field at its surface with a non-zero average. As the slip velocity originates from  local solute gradients, one natural way to create self-propulsion is to consider non isotropic particles 
with a reactive cap on an otherwise-inert surface. Tangential gradients in solute concentrations are then expected to be generated   between inert and active regions leading to slip velocity and locomotion. These so-called Janus particles are typically axisymmetric and have been the focus of most experimental, theoretical and numerical studies on autophoretic particles \citep{golestanian2005,golestanian2007,cordova2008,julicher2009,sabass2012,jiang2010}.

For such axisymmetric particles, the chemical properties of the surface are characterized by an activity $k=k(\mu)$ where $\mu=\cos\theta$, with $\theta$ the polar angle with respect to the axis of symmetry $\eb_z$ in spherical polar coordinates. The solute concentration and the flow field are also axisymmetric, and we denote $c=c(r,\mu)$, $\ub=u_r(r,\mu)\eb_r+u_\theta(r,\mu)\eb_\theta$. Consequently   the motion of the particle is a pure translation along $\eb_z$, $\Ub=U\eb_z$, with no rotation, $\Omegab=0$.

In this  setting, the Stokes flow problem can be solved explicitly using the squirmer formulation \citep{blake1971,michelin2011}. The flow velocity is completely determined by the streamfunction $\psi$, obtained as the superposition of orthogonal modes
\begin{equation}
\psi(r,\mu)=\sum_{n=1}^\infty\frac{2n+1}{n(n+1)}\alpha_n\psi_n(r)(1-\mu^2)L_n'(\mu),
\end{equation}
where $L_n(\mu)$ is the $n$th Legendre polynomial and
\begin{equation}
\psi_1(r)=\frac{1-r^3}{3r},\quad \psi_n(r)=\frac{1}{2}\left(\frac{1}{r^n}-\frac{1}{r^{n-2}}\right)\textrm{  for   }n\geq 2.
\end{equation}
The intensities of the squirming modes,  $\alpha_n$, are obtained through projection of the slip velocity $u_\theta(1,\mu)$ as
\begin{equation}
\alpha_n=\frac{1}{2}\int_{-1}^1\sqrt{1-\mu^2}L_n'(\mu)u_\theta(1,\mu)\dd\mu.\label{eq:alphan}
\end{equation}

The first squirming mode, $\alpha_1$, is the only one contributing to the swimming velocity of the particle, so that $\mathbf{U}=\alpha_1\eb_z$. The second squirming mode, $\alpha_2$, includes the slowest decaying contribution to the flow field, namely that of a symmetric force-dipole of intensity $\Sigma=10\pi\alpha_2$. The contribution of this particle to the bulk stress takes the form of a stresslet $\boldsymbol\Sigma=\Sigma \left(\mathbf{p}\mathbf{p}-\mathbf{I}/3\right)$ \citep{Batchelor1970S}. For $\Sigma>0$ ($\alpha_2>0$), this flow field is equivalent to a so-called {puller} swimmer swimming flagella first (such as the alga \emph{Chlamydomonas}), while $\Sigma<0$ ($\alpha_2<0$) corresponds to a {pusher} swimming body first (such as most flagellated bacteria and spermatozoa).

With this formalism, the flow field is completely characterized and determined by the intensities of the squirming modes  $\{\alpha_n\}_n$. Decomposing the surface reaction rate, $k(\mu)$, and solute distribution, $c(r,\mu)$, onto  Legendre polynomials,
\begin{equation}
k(\mu)=\sum_{p=0}^\infty k_pL_p(\mu),\qquad c(r,\mu)=\sum_{p=0}^\infty c_p(r)L_p(\mu),
\end{equation}
we can then rewrite the  phoretic problem, Eqs.~\eqref{eq:phoretic_01}--\eqref{eq:phoretic_02}, as a set of non-linearly coupled ODEs for the functions $c_p(r)$ ( $p\geq 0$), from which the characteristics of the flow fields can be retrieved. Specifically we obtain 
\begin{align}
\totd{}{r}\left(r^2\totd{c_p}{r}\right)&-p(p+1)c_p=\Pe\sum_{n=1}^\infty\sum_{m=0}^\infty\alpha_n\left[A_{mnp}\psi_n\totd{c_m}{r}+B_{mnp}\totd{\psi_n}{r}c_m\right]\label{eq:phoretic_11},\\
&c_p(\infty)=0,\label{eq:phoretic_12}\\
\totd{c_p}{r}(1)&=k_p+\Da\sum_{m=0}^\infty\sum_{n=0}^\infty\frac{A_{mnp}k_n}{2n+1}c_m(1)\label{eq:phoretic_13},\\
\alpha_n&=-\frac{n(n+1)M}{2n+1}c_n(1)\quad\textrm{for}\quad n\geq 1,\label{eq:phoretic_14}
\end{align}
where the third-order tensors $A_{mnp}$ and $B_{mnp}$ are defined from the Legendre polynomials as \citep{michelin2011}
\begin{align}
A_{mnp}=&\frac{(2p+1)(2n+1)}{2}\int_{-1}^1L_m(\mu)L_n(\mu)L_p(\mu)\dd\mu,\label{eq:defA}\\
 B_{mnp}=&\frac{(2p+1)(2n+1)}{2n(n+1)}\int_{-1}^1(1-\mu^2)L_m'(\mu)L_n'(\mu)L_p(\mu)\dd\mu.\label{eq:defB}
\end{align}

\subsection{Janus particles}
\begin{figure}
\begin{center}
\includegraphics[width=.65\textwidth]{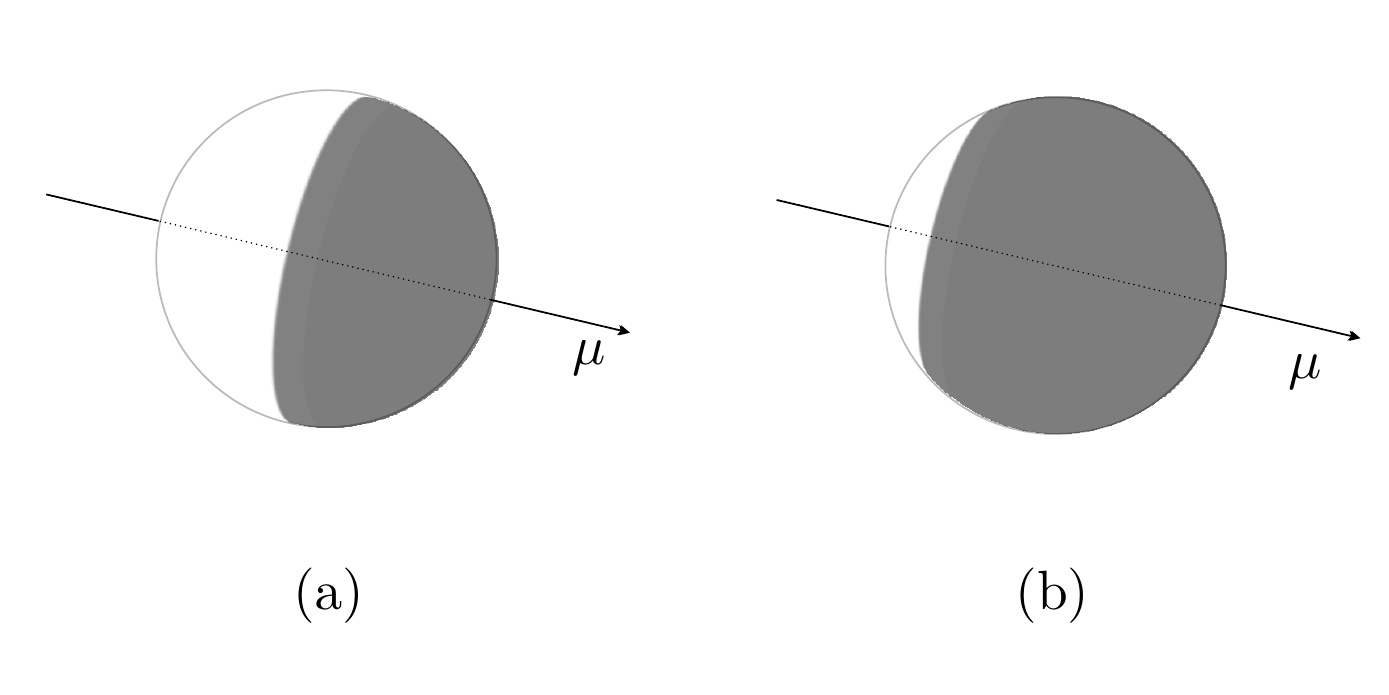}
\caption{Janus particles A and B considered in our computations: (a) Particle A is a hemispheric Janus particle with one half inert (white) and one half active (grey). (b) Particle B is a non-symmetric Janus swimmer with a large reactive pole (grey) and small inert area (white).}\label{fig:Janus}
\end{center}
\end{figure}

The Janus particles considered here consist of a reactive cap at one pole of the sphere while the rest of the particle is inert. A variety of  Janus particles are considered here  and they differ by the ratio of their inert to active surface area (Fig.~\ref{fig:Janus}). The chemical activity distribution is given by $ k(\mu)=1_{\{\mu>\mu_c\}}$, where 
$\mu_c$ denotes the angular size of the active region ($-1\leq \mu_c\leq 1$). We assume for simplicity that the phoretic  mobility is uniform ($M=\pm 1$). That assumption however does not impact our main results, which may easily be generalized to Janus mobility distributions such that $M(\mu)=\pm k(\mu)$ as shown in  Appendix~\ref{sec:nonunif_mob}. For the choice $ k(\mu)=1_{\{\mu>\mu_c\}}$, the spectral coefficients $k_n$ of the activity distribution can be obtained by projection of $k(\mu)$ onto the Legendre polynomials and we obtain
\begin{equation}
k_0=\frac{1-\mu_c}{2},\, \textrm{  and  } k_n=\frac{1}{2}\left[L_{n-1}(\mu_c)-L_{n+1}(\mu_c)\right]\textrm{   for   } n\geq 1.\label{eq:janusmodes}
\end{equation}

In the limit where both advective and reactive effects can be neglected ($\Pe=\Da=0$), the diffusive problem for $c$, Eqs.~\eqref{eq:phoretic_11}--\eqref{eq:phoretic_13}, can be solved analytically along each azimuthal component and we obtain
\begin{equation}\label{eq:ref_sol1}
c_p(r)=-\frac{k_p}{(p+1)r^{p+1}}\cdot
\end{equation}
Then, Eq.~\eqref{eq:phoretic_14} provides the squirming mode intensities
\begin{equation}\label{eq:ref_sol2}
\alpha_p=\frac{p\,k_p M}{2p+1},
\end{equation}
from which the entire flow field can be computed; in particular the reference swimming velocity and stresslet, thereafter referred to as $U_0$ and $\Sigma_0$, are obtained as
\begin{equation}
U_0=\frac{k_1M}{3}=\frac{M}{4}(1-\mu_c^2), \quad \Sigma_0=4\pi Mk_2=5\pi M\mu_c(1-\mu_c^2).
\end{equation}

For our computations in  Section~\ref{sec:results}, we specifically focus on two such Janus particles denoted Particles A and B (Fig.~\ref{fig:Janus}) of uniform mobility $M=\pm 1$, this effectively amounting to four different configurations, or two pairs.

\begin{figure}
\begin{center}
\includegraphics[width=.8\textwidth]{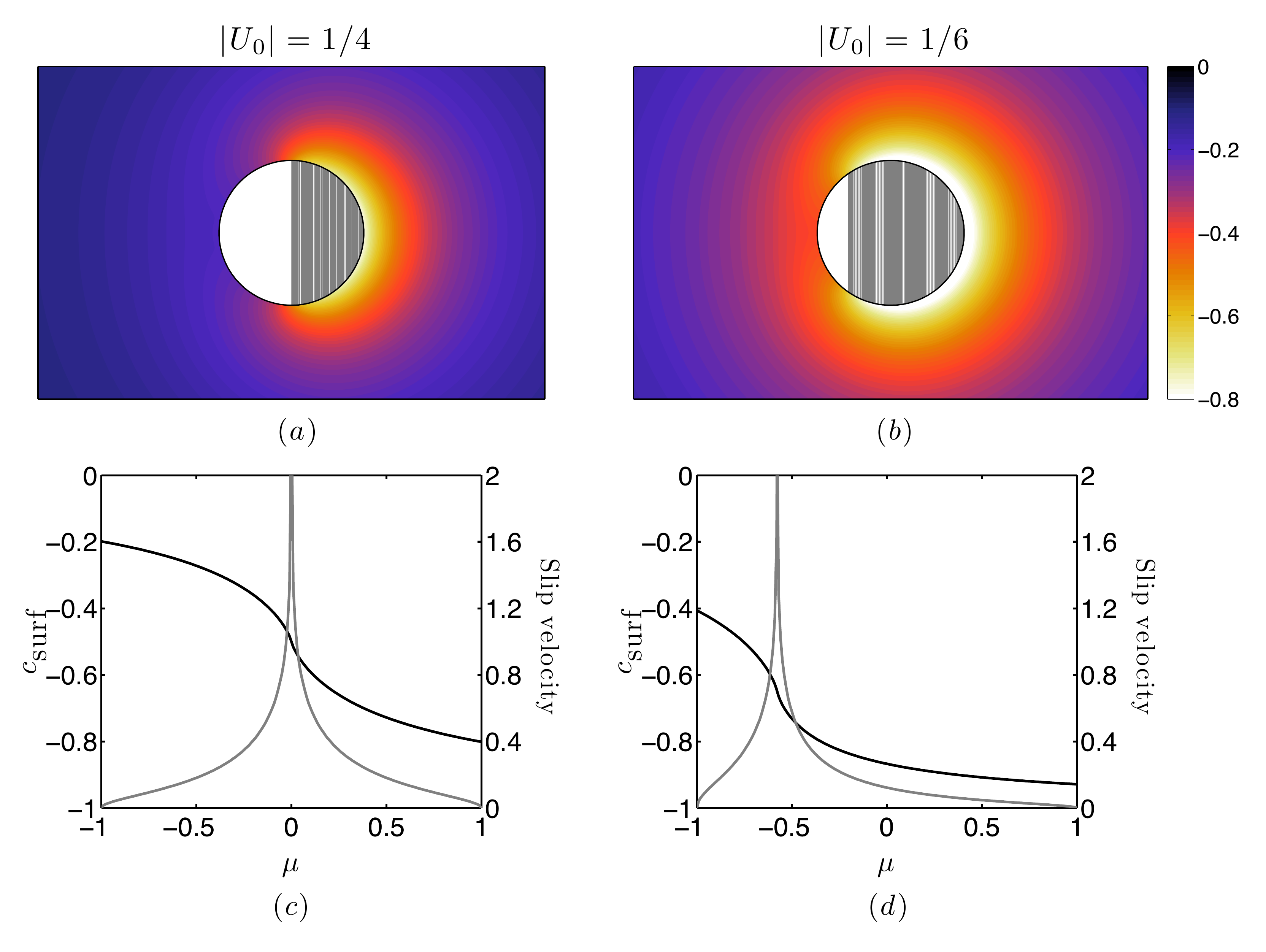}
\caption{(Colour online) Top: Relative solute concentration distribution, $c$, in the reference configuration with no advective or reactive effects ($\Pe=\Da=0$) for (a) Janus particle A and (b) Janus particle B. The reactive cap is shown in grey and the inert portion in white.  For $M=1$ (resp.~$M=-1$), the swimming velocity is oriented to the right (resp.~left). 
Bottom: Surface solute concentration (black) and slip velocity (grey) along the surface in the case of positive mobility ($M=1$) for (c) particle A and (d)  particle B.}\label{fig:concentration_reference}
\end{center}
\end{figure}

Particle A is a hemispheric (symmetric) Janus swimmer ($\mu_c=0$) with one half chemically-active  and the other half inert. The corresponding spectral coefficients $k_n$ are computed as
\begin{equation}
k^A_0=\frac{1}{2}, \quad k^A_{2q}=0, \textrm{   and  }k^A_{2q-1}=(-1)^{q+1}\frac{4q-1}{4q-2}\frac{(2q)!}{\left[2^qq!\right]^2}\quad \textrm{for   }q\geq 1.
\end{equation}
In particular, $k^A_1=3/4$ and $k^A_2=0$, so that $U_0^A=M/4$ and $\Sigma_0^A=0$. This is the particle with maximum swimming velocity in the limit where both advective and reactive effects are negligible ($\Pe=\Da=0$). The sharpest concentration gradients are located near the equator for particle A, resulting in the largest slip velocities located on an extended surface and oriented mostly horizontally (as illustrated in Fig.~\ref{fig:concentration_reference}a and c). In contrast, particle A has no stresslet in the $\Pe=\Da=0$ limit and its far-field signature has a faster decay and is dominated by a source dipole and a force quadrupole.

Particle B is a non-symmetric Janus swimmer with $\mu_c=-1/\sqrt{3}$. It consists of a larger active cap and small inert portion. Using Eq.~\eqref{eq:janusmodes}, one obtains that $k_1^B=1/2$ and $k_2^B=5/6\sqrt{3}$, so that for particle B, we get a smaller swimming velocity, $U_0^B=M/6$,  and a finite stresslet,  $\Sigma_0^B=-10\pi M/3\sqrt{3}$. The front between reactive and inert regions is located closer to the pole, and thus involves a smaller share of the particle  surface creating slip velocities inclined away from the direction of motion and therefore   a smaller swimming speed (see Fig.~\ref{fig:concentration_reference}b and d).  On the other hand, particle B corresponds to a maximum stresslet intensity for $\Pe=\Da=0$, and as such is one of the  Janus particle inducing the largest far-field hydrodynamic interactions with other particles. 

Note that  for both particles, the chemical reaction at the surface results in a reduction of the solute concentration near the reactive pole ($\mu=1$). For a slip velocity oriented along (resp.~against) the surface gradient, $M=1$ (resp.~$M=-1$), the slip velocity is oriented from the reactive to the inert pole (resp.~from the inert pole to the reactive pole) resulting in a positive (resp.~negative) swimming velocity.

\subsection{Numerical solution}
For finite values of $\Da$ and $\Pe$, the phoretic problem, Eqs.~\eqref{eq:phoretic_11}--\eqref{eq:phoretic_14}, is solved numerically for each particle. The different azimuthal modes of the solute distribution, $c_p(r)$, are discretized on a stretched radial grid \citep{michelin2011,michelin2013b}, and an iterative process is followed:
\begin{enumerate}\renewcommand{\theenumi}{\roman{enumi}}
\item{For an initial guess of the flow, determined by an initial guess of $\{\alpha^i_n\}_n$, the linear advection diffusion problem, Eqs.~\eqref{eq:phoretic_11}--\eqref{eq:phoretic_13}, is solved directly for $\{c_p(r)\}_p$ \citep{michelin2011}.}
\item{Using the solution of this advection-diffusion problem, Eq.~\eqref{eq:phoretic_14} is used to obtain an updated estimate of the squirming mode intensities, $\{\alpha^f_n\}_n$.}
\item{Broyden's method is used to solve iteratively the non-linear system $\mathbf{F}(\alphab)=\alphab^f-\alphab^i$: knowing an estimate $\alphab^n$ of the solution and an estimate of the inverse of the Jacobian matrix $\mathbf{J}^{-1}_n=\left[\nabla_{\alphab} \mathbf{F}(\alphab^n)\right]^{-1}$, a new estimate for both quantities is obtained as \citep{broyden1965}
\begin{align}
\alphab^{n+1}&=\alphab^n-\mathbf{J}^{-1}_n\cdot\alphab^n,\\
\mathbf{J}^{-1}_{n+1}&=\mathbf{J}^{-1}_{n}+\frac{(\Delta\alphab-\mathbf{J}^{-1}_{n}\Delta\mathbf{F})\cdot(\Delta\alphab^T\cdot\mathbf{J}^{-1}_{n})}{\Delta\alphab^T\cdot\mathbf{J}^{-1}_{n}\cdot\Delta\alphab},
\end{align}
where $\alphab^T$ is the transpose of the column vector $\alphab$, $\Delta\alphab=\alphab^{n+1}-\alphab^n$, and $\Delta\mathbf{F}=\mathbf{F}(\alphab^{n+1})-\mathbf{F}(\alphab^n)$. The iteration is initiated using either a previous computation or the reference solution ($\Da=\Pe=0$), in which case the Jacobian matrix must be computed numerically for the initial step. 
}
\end{enumerate}

The number and position of the points on the radial grid, as well as the number of azimuthal modes used for the solute concentration distribution, are adjusted to the value of the P\'eclet number \citep{magar2003,michelin2011}. Typical computations for moderate $\Pe$ include 120 azimuthal modes and 150 radial points. Truncation must also be introduced in the number of squirming modes used and the number of azimuthal components retained for $k(\mu)$. In the results presented below, typically $n_\alpha=8$ squirming modes were used as well as $n_k=12$ modes for the surface activity. These two parameters critically impact computational cost. Convergence tests performed showed that for $\Pe$ less than 100, the swimming velocity was only marginally impacted (less than $0.5\%$) when doubling $n_\alpha$ or $n_k$.

\section{Advective and reactive effects on the self-propulsion of Janus particles}
\label{sec:results}
Using the model and formalism presented in Section~\ref{sec:model}, we now investigate the effect of advection ($\Pe$) and reaction ($\Da$) on the self-propulsion of Janus phoretic particles. The asymmetry of the particles   ensure that self-propulsion is achieved for all values of $\Pe$ and $\Da$ even in the purely diffusive regime $\Pe=\Da=0$. Note that self-propulsion can also be achieved by isotropic or symmetric particles as an instability in the  nonlinear advective coupling of solute dynamics to the phoretic flow around the particle \citep{michelin2013c}. In that case, and in certain conditions,  a critical $\Pe$ exists above which symmetry-breaking leads to propulsion.

\subsection{Advective effect for the fixed-flux limit ($\Da=0$)}

\begin{figure}
\begin{center}
\begin{tabular}{cc}
\includegraphics[height=.44\textwidth]{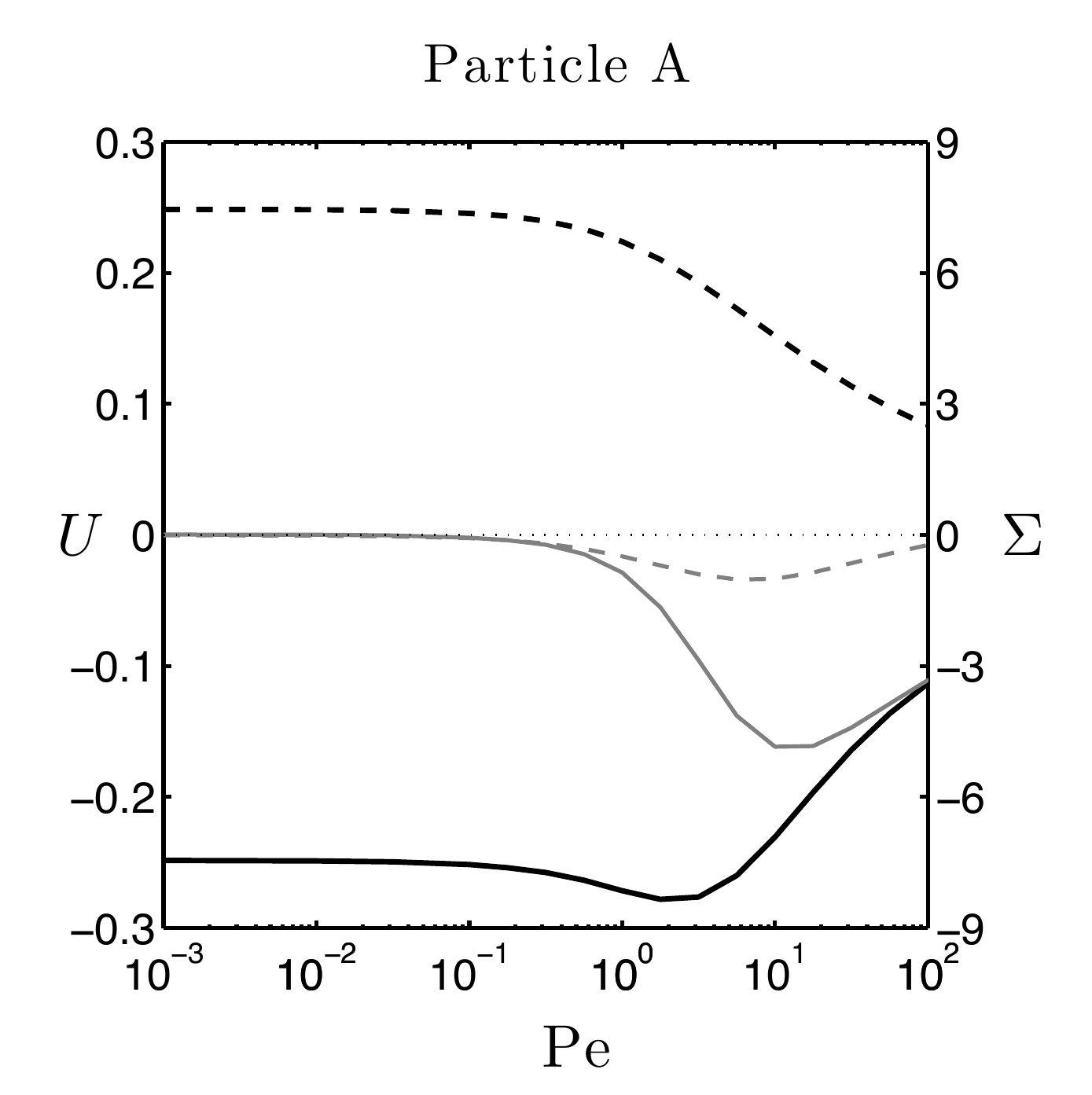} &
\includegraphics[height=.44\textwidth]{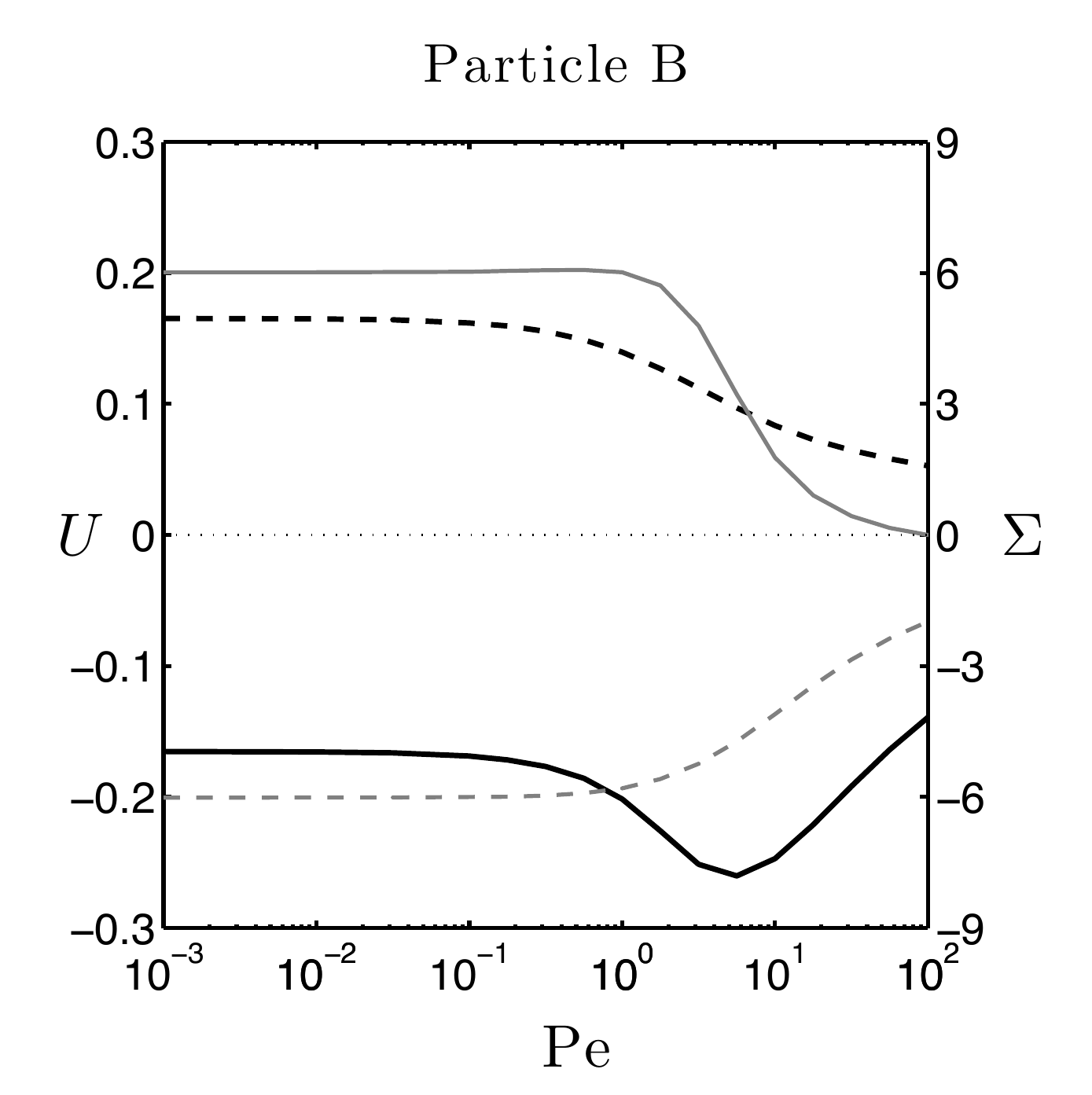}
\end{tabular}
\caption{Dependence of the swimming velocity ($U$, black) and stresslet magnitude ($\Sigma$, grey)  on the value of $\Pe$  for particle A (left) and particle B (right) in the fixed-flux limit ($\Da=0$). For each case, results are shown for negative mobility $M=-1$ (solid) and positive mobility $M=1$ (dashed).}\label{fig:Peeffect}
\end{center}
\end{figure}

We start by considering the effect of advection on phoretic locomotion. When $\Pe$ is increased, the solute concentration distribution around the particle is modified due to the advection of the solute by the flow resulting from the phoretic slip velocity. As a result, local concentration gradients and the slip velocity distribution are also impacted, and changes in the swimming velocity occur. For both Janus particles A and B, and for both values of the mobility ($M=\pm 1$), Figure~\ref{fig:Peeffect} shows  the dependence  of the swimming velocity, $U$, and the  stresslet intensity, $\Sigma$,  on the P\'eclet number,  $\Pe$,  in the absence of  reactive effects  ($\Da=0$). This situation  corresponds therefore to a fixed-flux solute absorption at the surface. The case of a fixed-flux solute release is obtained directly by changing $M$ into $-M$.  

At large values of  $\Pe$, when advection dominates over solute diffusion, the magnitude of the swimming velocity is seen to always decrease for both particles. Analysis of the numerical data in Fig.~\ref{fig:Peeffect} suggests that both $|U|,|\Sigma|\sim\Pe^{-1/3}$ at large $\Pe$. This scaling is consistent with that suggested by Ref.~\citep{julicher2009} and can be recovered from dimensional analysis \change{as follows. At large $\Pe$, the solute distribution resulting from the advection/diffusion problem is characterized by a boundary layer. The boundary layer thickness $\delta$ is the typical length scale associated with radial gradients of the solute concentration, while the typical length scale associated with tangential gradients remains $O(1)$ (i.e. the radius of the spherical particle). The boundary layer thickness $\delta$ is then obtained by balancing normal diffusive flux ($\sim \mathcal{C}/\delta^2$) with tangential advection ($\sim\Pe\,\mathcal{U}\mathcal{C}$) near the surface, leading to} $\delta\sim(\Pe\,\mathcal{U})^{-1/2}$, with $\mathcal{U}$ the typical slip velocity \citep{michelin2011}. When $\Da=0$, the normal diffusive flux is fixed and $O(1)$, therefore $\mathcal{C}\sim\delta$ is the typical scale of variation of the solute concentration at the surface. Finally, the definition of the phoretic slip velocity \change{ in Eq.~\eqref{eq:slip}} imposes that $\mathcal{C}\sim\mathcal{U}$. \change{Combining these three scaling arguments leads to}  $\mathcal{U},\mathcal{C},\delta\sim\Pe^{-1/3}$, and the same dependence \change{with \Pe} is recovered for the swimming velocity and stresslet intensity.

Although the autophoretic velocity decreases for all particles with the same scaling at large $\Pe$, their finite-$\Pe$ evolutions  strongly differ depending on the sign of the mobility. Particles A and B with positive mobility ($M=1$) swim in the direction of their active pole (i.e. to the right in in Figs.~\ref{fig:Janus} and  \ref{fig:concentration_reference}), and their velocity decreases {monotonically} in magnitude for all $\Pe$. In contrast, particles with negative mobility ($M=-1$) swim toward their inert pole (i.e. to the left), and their velocity varies {non-monotonically} reaching a maximum magnitude around $\Pe\approx 2$, before decreasing as $\Pe^{-1/3}$. The existence of this velocity maximum is a new and notable result, which is not restricted to this type of swimmers. Indeed,  several different particle activity distributions were tested, leading to the same result. For any given activity distribution $k(\mu)$, if the velocity magnitude of a particle of mobility $M$ decreases monotonically with $\Pe$, then for the particle with the same activity and opposite mobility $-M$, the velocity magnitude shows a peak in magnitude at intermediate values of  $\Pe$. For Janus particles, our simulations indicate in fact that  regardless of the coverage of the active cap (i.e. for all $\mu_c$), particles with negative mobility experienced a velocity peak at intermediate $\Pe$, while particles with positive mobility exhibited a velocity decreasing monotonically with $\Pe$.

\begin{figure}
\begin{center}
\includegraphics[width=.8\textwidth]{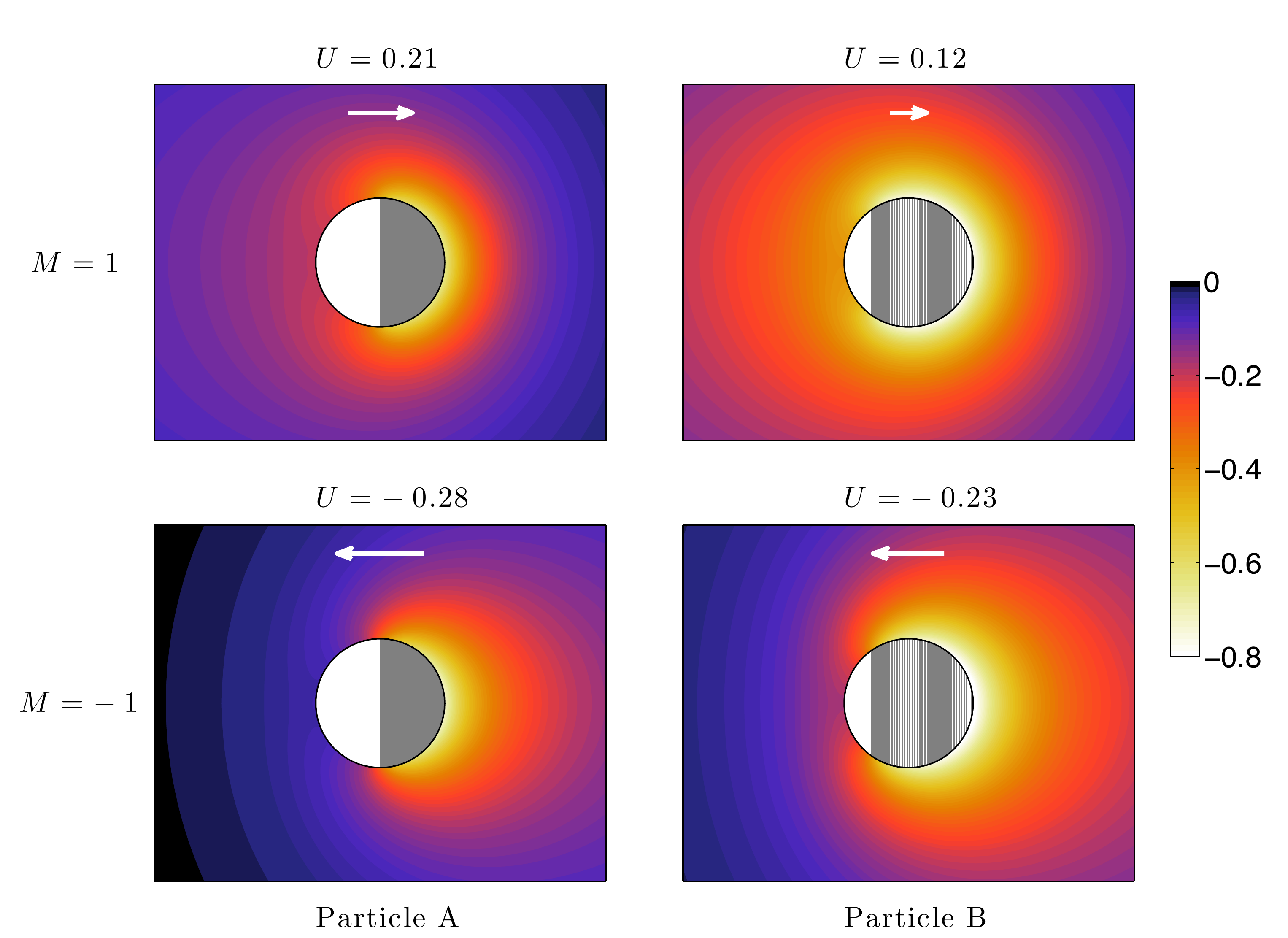}
\caption{(Colour online) Relative concentration distribution $c$ around phoretic  particle A (left) and  particle B (right), for $\Pe=2$ and $\Da=0$. The swimming velocity of each particle is indicated by a white arrow. The reactive part of the surface is shown in grey.}\label{fig:concentration_Pe2Da0}
\end{center}
\end{figure}

\begin{figure}
\begin{center}
\includegraphics[width=.8\textwidth]{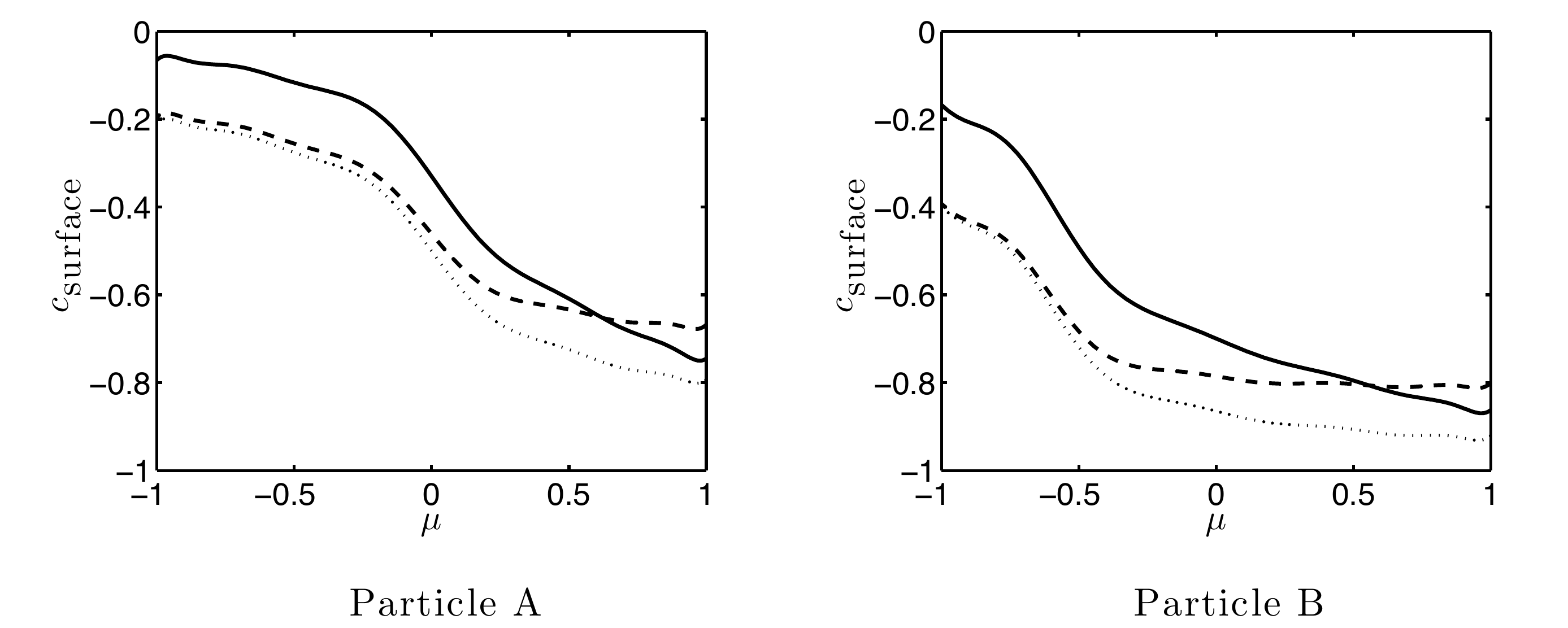}
\caption{Surface concentration distribution for particles A (left) and B (right) for $\Pe=2$ and $\Da=0$, in the case of positive mobility ($M=1$, dashed) and negative mobility ($M=-1$, solid). The distribution in the reference configuration, $\Pe=\Da=0$, is shown for reference as a dotted line.}\label{fig:surf_conc_Pe2}
\end{center}
\end{figure}

\begin{figure}
\begin{center}
\includegraphics[width=.8\textwidth]{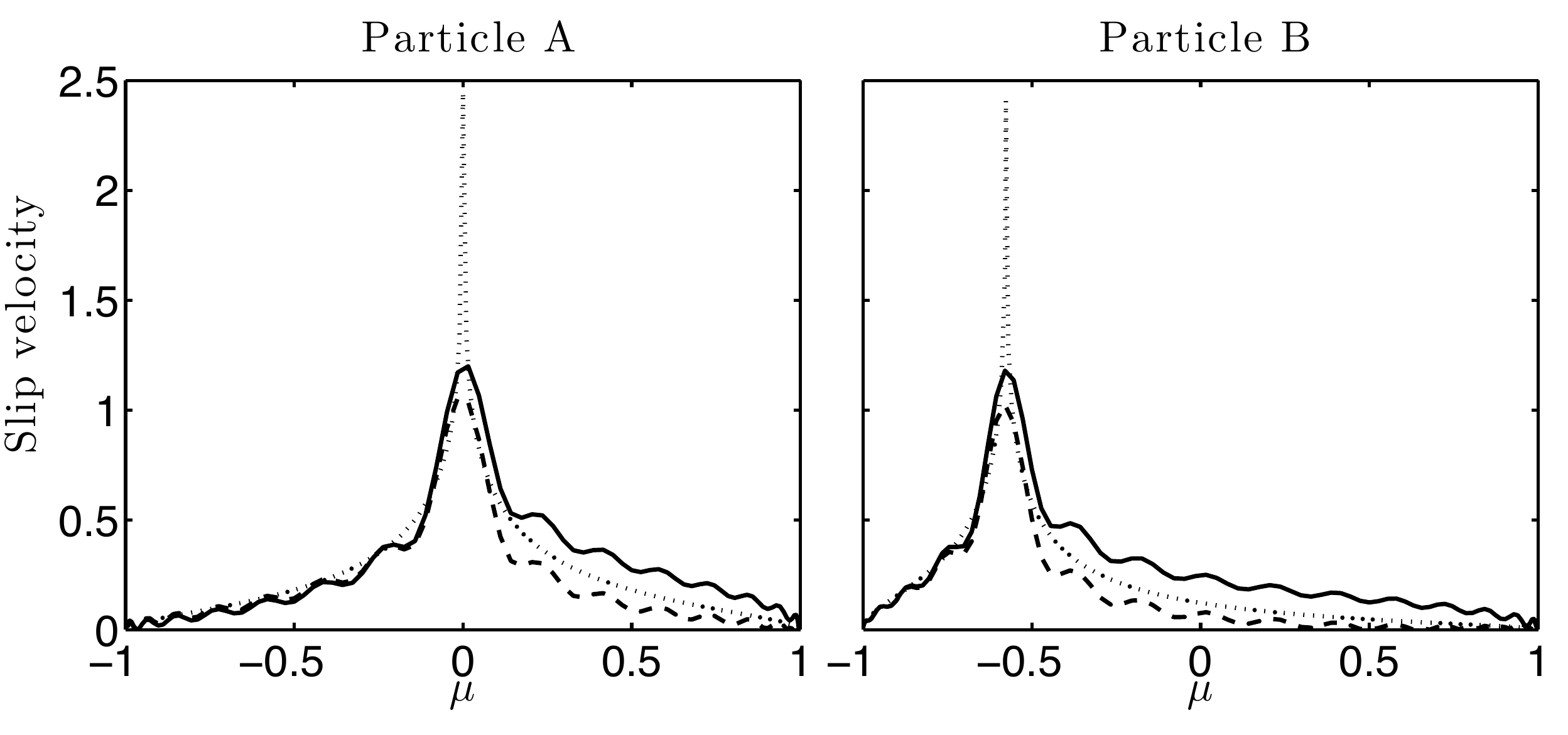}
\caption{\change{Surface slip velocity distribution for particles A (left) and B (right) for $\Pe=2$ and $\Da=0$, in the case of positive mobility ($M=1$, dashed) and negative mobility ($M=-1$, solid). The slip velocity in the reference configuration, $\Pe=\Da=0$, is reported for reference as a dotted line. Note that due to the computational cost, only 32 modes were used to describe the slip velocity for $\Pe=2$ while the slip velocity in the reference configuration can be obtained analytically from Eqs.~\eqref{eq:ref_sol1}--\eqref{eq:ref_sol2}.}}\label{fig:slipvel_Pe2}
\end{center}
\end{figure}

For Janus particles, this difference in behaviour depending on the sign of the mobility can be qualitatively understood by comparing the solute concentration distribution around the particles (see Figs.~\ref{fig:concentration_Pe2Da0} and \ref{fig:surf_conc_Pe2}) with the reference situation at $\Pe=0$ (Fig.~\ref{fig:concentration_reference}). For all particles and all $\Pe$, the chemical reaction near the active pole ($\mu=1$) results in a depletion zone of the solute concentration in that region.  When the particle swims toward the inert pole ($M=-1$, bottom plots in Fig.~\ref{fig:concentration_Pe2Da0}), advection of the fluid along the surface tends to concentrate this depleted region in a narrower region in the wake of the particle. Most importantly, advection brings in the vicinity of the front pole (here the inert one) fluid with higher solute content. Both effects exacerbate the concentration contrast between the fore and aft poles and the solute gradients along the surface, resulting in an increase in the slip and swimming velocity magnitudes when $\Pe$ is increased (Figs.~\ref{fig:surf_conc_Pe2} \change{and \ref{fig:slipvel_Pe2}}). \change{Note that the increase in slip velocity is limited to the reactive region. The slip velocity in the inert region remains roughly identical to the reference configuration. For absorbing particles with negative mobility (or equivalently for emitting particles with positive mobility), the advection of the solute by the phoretic flows introduces a positive feedback on the swimming velocity, that is similar to the one identified by Ref.~\citep{michelin2013c} on isotropic particles and responsible, in that case, for symmetry breaking and propulsion. The similarity between these two problems is further discussed in Section~\ref{sec:janus_general}.}

 In contrast, when the particle swims toward the reactive pole ($M=1$, top plots in Fig.~\ref{fig:concentration_Pe2Da0}), the advection of richer fluid toward the reactive pole tends to increase the concentration in this depleted region, which is also spread on a larger part of the particle by the tangential advection along the surface. Both effects tend this time to reduce the concentration contrast between fore and aft poles resulting in a reduction of the slip and swimming velocity magnitudes (Fig.~\ref{fig:slipvel_Pe2}): \change{solute advection by phoretic flows leads in this case to a negative feedback}. 
 
 When $\Pe\gg 1$, however, advection tends to homogenize the solute concentration near the boundary except in a narrow wake region: regardless of the sign of mobility and of the swimming direction, advective effects eventually penalize phoretic propulsion.

This difference of behaviour (i.e. existence of an extremum vs.~monotonic decrease) is also observed for the stresslet, $\Sigma$, when $\Sigma_0\neq 0$ (particle B). However, the magnitude of the peak differs only marginally from the stresslet amplitude in the reference configuration ($|(\Sigma_\textrm{max}-\Sigma_0)/\Sigma_0|\approx 0.9\%$) so that it is barely visible on Figure~\ref{fig:Peeffect}. For Janus particle A, $\Sigma_0=0$ when $\Pe=0$ by symmetry, and the stresslet magnitude $\Sigma$ is always negative for $\Pe>0$ and $\Da=0$.  The impact of the  phoretic particle  on the far-field flow is that of a pusher swimmer, similar to that of most flagellated bacteria. For both particles, the stresslet magnitude is maximum for $\Pe=O(10)$. These results are confirmed and extended to arbitrary Janus particles in Section~\ref{sec:janus_general}.

\subsection{Reactive effects in the diffusive limit ($\Pe=0$)}

We now consider the effect of reaction kinetics on the swimming velocity when advective effects are neglected ($\Pe=0$). At finite values of  
$\Da$, the rate of solute absorption becomes dependent on the local solute concentration. The reaction at the surface is   fast enough for diffusion to be unable to maintain a relatively homogeneous background concentration of solute around the particles. In other words, the concentration changes induced by surface reaction are now of similar magnitude to the background/far-field concentration. As a result, the reaction rate will be reduced in regions where the solute concentration is lower, in particular near the active pole.

\begin{figure}
\begin{center}
\begin{tabular}{cc}
\includegraphics[height=.42\textwidth]{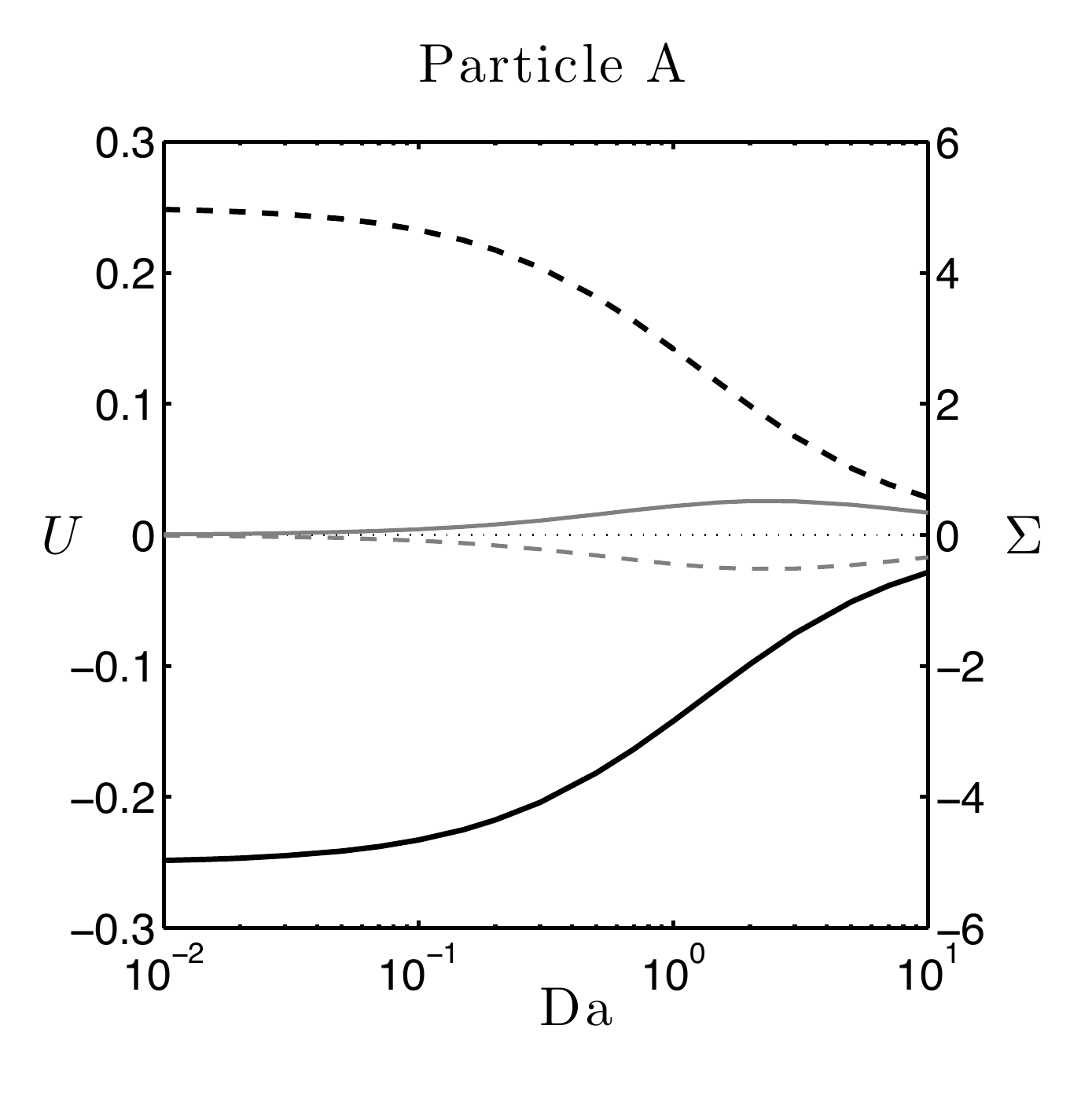} &
\includegraphics[height=.42\textwidth]{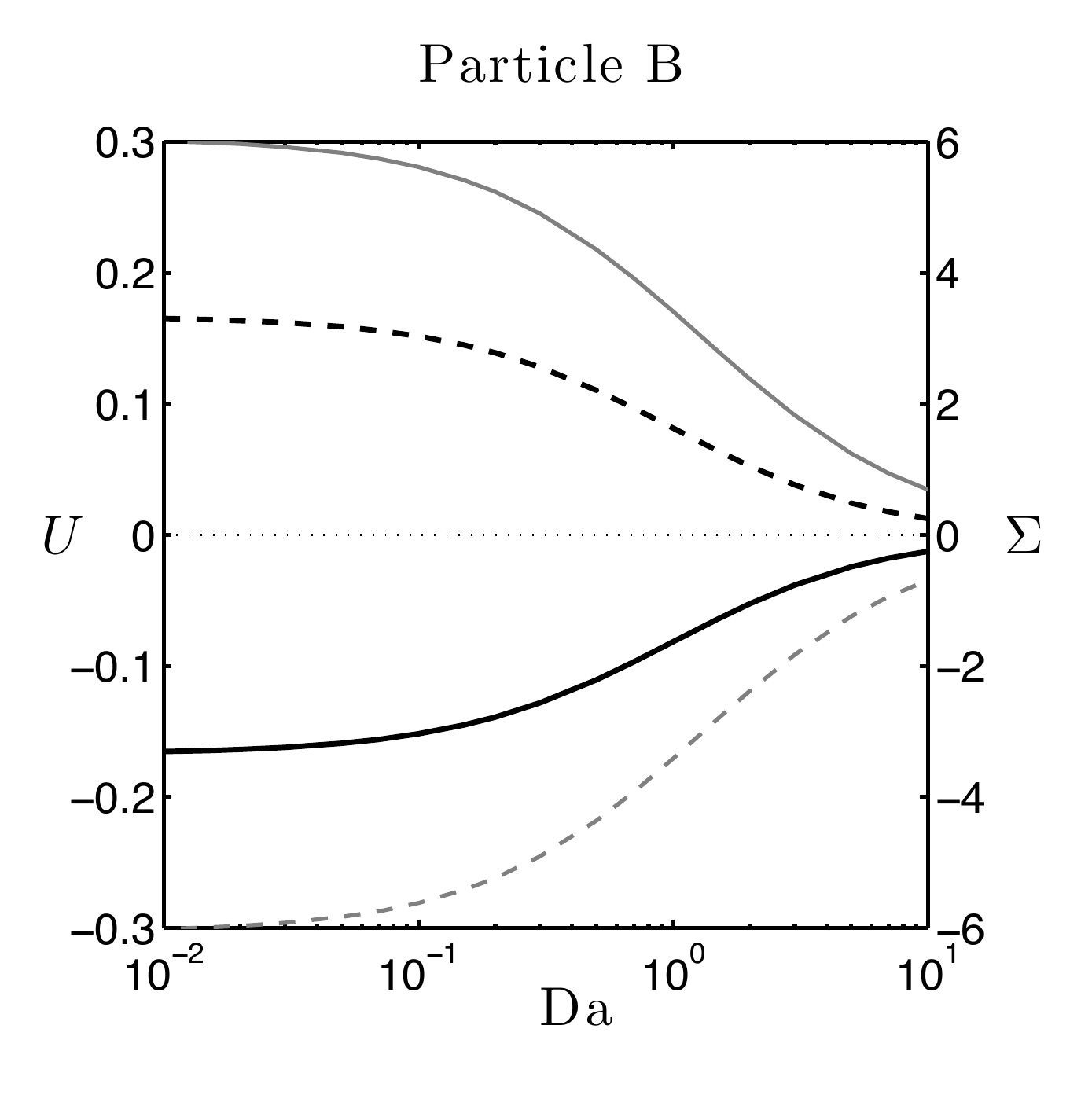}
\end{tabular}
\caption{Dependence  of the swimming velocity ($U$, black) and stresslet magnitude ($\Sigma$, grey)  with $\Da$ for particle A (left) and particle B (right) in the diffusive limit ($\Pe=0$). Results are obtained for both negative mobility $M=-1$ (solid) and positive mobility $M=1$ (dashed).}\label{fig:Daeffect}
\end{center}
\end{figure}

\begin{figure}
\begin{center}
\includegraphics[width=.9\textwidth]{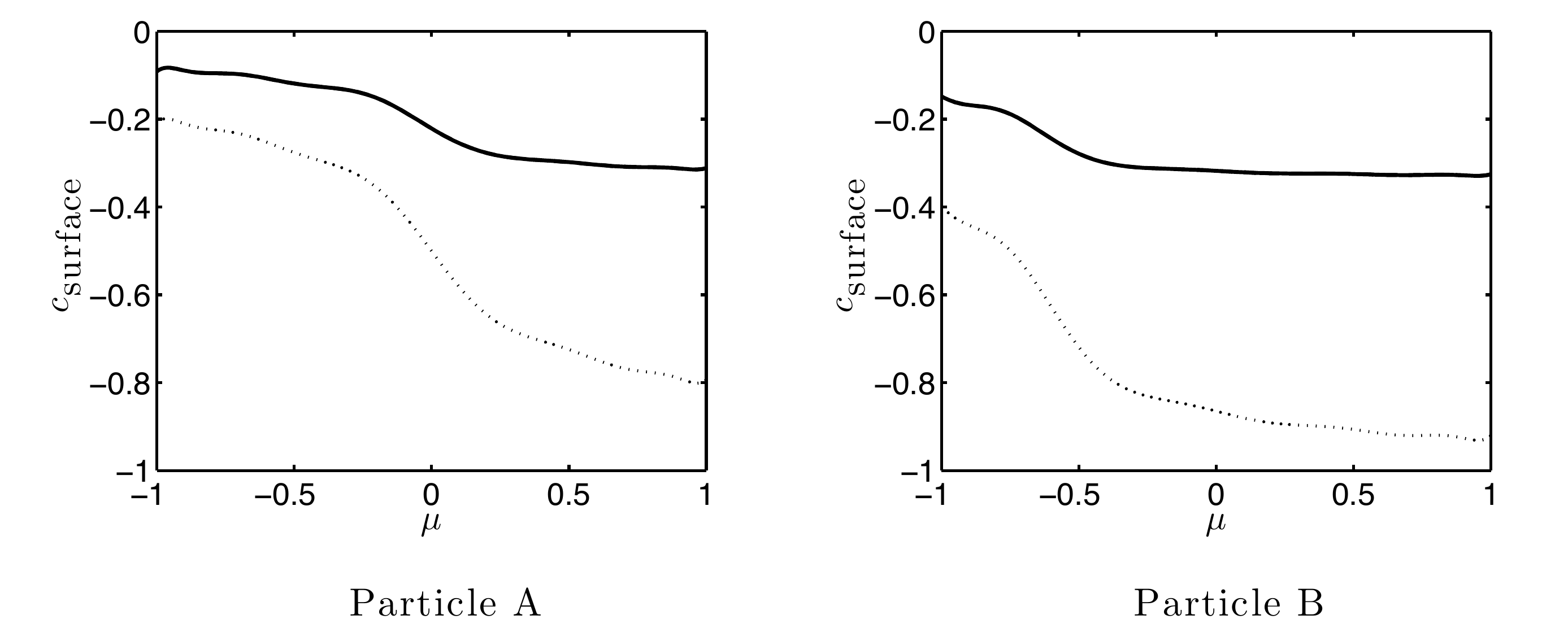}
\caption{Surface concentration distribution for particles A (left) and B (right) for $\Da=2$ and $\Pe=0$ (solid). Because advective effects are neglected, the concentration distribution is the same for both $M=1$ and $M=-1$. The distribution in the reference configuration, $\Pe=\Da=0$, is shown for reference as a dotted line.  }\label{fig:surf_conc_Da2}
\end{center}
\end{figure}

The dependence of the velocity and stresslet intensity with $\Da$ is shown on Fig.~\ref{fig:Daeffect}. In contrast with the evolution of those quantities with solute advection, we observe a strong symmetry between the cases of positive and negative mobility: in the absence of any advective effects, solute concentration is determined purely by diffusion and has the same distribution regardless of the mobility of the particle. Particles of opposite mobilities have exactly opposite slip velocity distributions. Also, no peak in the velocity magnitude can be observed. Instead, the swimming velocity monotonically decreases with $\Da$ and tends to zero in the limit $\Da\gg 1$. This is consistent with the comment above on the role of diffusion vs.~reaction. Indeed, for larger values of $\Da$, the reaction leading to the absorption of the solute is slowed down near the active pole (where the solute concentration is lowest), as illustrated in  Fig.~\ref{fig:surf_conc_Da2}. As a result the tangential concentration gradients are reduced and so are the slip and swimming velocities, regardless of the value of the mobility $M$. In the limit of $\Da\gg 1$, a decrease of the swimming velocity as $\Da^{-1}$ is observed. In that limit, the diffusion timescale is infinite, leading to a complete depletion of the most reactive regions. The perturbations to the solute concentration on the surface of the particle scale thus as $c\sim -\Da.^{-1}$, resulting in   a similar scaling for the slip and swimming velocities.

\subsection{Finite P\'eclet and Damk\"ohler numbers}

\begin{figure}
\begin{center}
\includegraphics[width=.85\textwidth]{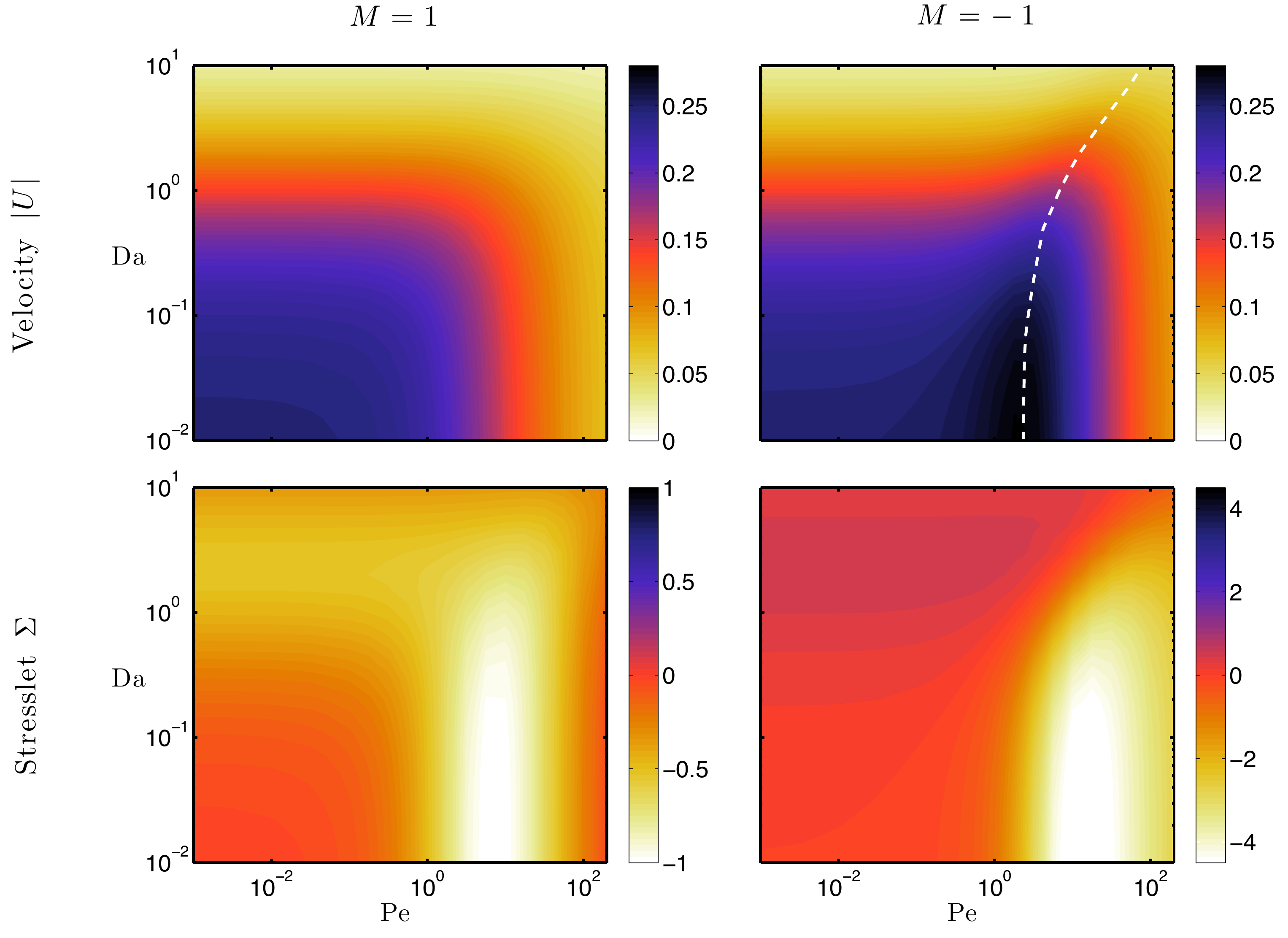}
\caption{(Colour online) Dependence of the magnitude of the propulsion velocity ($|U|$, top) and the stresslet intensity ($\Sigma$, bottom) with $\Pe$ and $\Da$ for the Janus particle A with $M=1$ (left) or $M=-1$ (right). The white dashed line on the top-right figure indicates the evolution with $\Da$ of the optimal $\Pe$ for which the velocity magnitude is maximum at fixed $\Da$. }\label{fig:USigma_janus}
\end{center}
\end{figure}

We show in Fig.~\ref{fig:USigma_janus} the dependence  of both $U$ and $\Sigma$ with finite values of  $\Pe$ and $\Da$, and confirm the results obtained in the limits $\Pe=0$ and $\Da=0$. Regardless of the value of the P\'eclet number, the swimming velocity magnitude is observed to monotonically decrease with $\Da$. Further,  regardless of the value of the Damk\"ohler number, the dependence of the swimming velocity with $\Pe$ is different for particles with positive and negative mobility. The swimming velocity of particles with positive mobility decreases monotonically while that of particles with negative mobility shows a maximum value for an intermediate $\Pe$. Note also that the optimal $\Pe$ leading to this velocity maximum appears to be an increasing function of $\Da$.

\section{Sensitivity of arbitrary Janus particles to advective and reactive effects}
\label{sec:janus_general}
In the previous section, the effects of advection ($\Pe$) and reaction ($\Da$) on the swimming velocity and stresslet intensity of two particular Janus particles were investigated computationally. In particular, it was shown that  (i) advection may increase the magnitude of the self-propulsion, when the particle is swimming toward its inert pole at $\Pe=0$, (ii)  reactive effects ($\Da>0$) always penalize self-propulsion and  (iii) advective effects create a negative stresslet on particle A (for which $\Sigma_0=0$), resulting in a pusher swimmer. In this section, we first  confirm these results and extend them to more general surface coverage using asymptotic analysis in the limit $(\Pe,\Da)\ll 1$. In particular, the sensitivity of the swimming velocity and stresslet to advective and reactive effects are mathematically determined by analytical calculations of four partial derivatives. For arbitrary $\Pe$, the evolution of the swimming velocity and stresslet intensity for arbitrary Janus particles is then  addressed numerically. 

\subsection{Asymptotic analysis for the autophoretic velocity}
\label{sec:asymptotics}
From Eqs.~\eqref{eq:phoretic_11}--\eqref{eq:phoretic_14}, we note that advective effects ($\Pe$) and reactive effects ($\Da$) are responsible for the coupling of the different azimuthal modes. In the limit where $\Da=\Pe=0$, the different modes decouple and the solution is obtained explicitly as
\begin{equation}
\bar{c}_p(r)=-\frac{k_p}{(p+1)r^{p+1}},\qquad \bar{\alpha}_p=\frac{pk_p M}{2p+1}\cdot\label{eq:refsol}
\end{equation}
Defining the corrections $c_p'=c_p-\bar{c}_p$ and $\alpha_p'=\alpha_p-\bar\alpha_p$ to this reference solution, using
\begin{align}
A_{mn1}=\frac{3(n+1)}{2n+3}\delta_{m,n+1}+\frac{3n}{2n-1}\delta_{m,n-1},\\
B_{mn1}=\frac{3(n+2)}{2n+3}\delta_{m,n+1}+\frac{3(n-1)}{2n-1}\delta_{m,n-1},
\end{align}
and keeping only the linear terms in the correction quantities, Eqs.~\eqref{eq:phoretic_11}--\eqref{eq:phoretic_13} become for the swimming mode ($p=1)$
\begin{align}
\totd{}{r}\left(r^2\totd{c'_1}{r}\right)-2c'_1=\Pe\,\bar\alpha_1&\left[k_0\left(\frac{1}{r^3}-1\right)+\frac{3k_2}{5r^5}\right]\nonumber\\
+\frac{3\Pe}{2}&\sum_{n=2}^\infty\bar\alpha_n\left[k_{n+1}\left(\frac{2n+1}{(2n+3)r^{2n+3}}-\frac{2n-1}{(2n+3)r^{2n+1}}\right)\right.\nonumber\\
&\left.+k_{n-1}\left(\frac{1}{r^{2n+1}}-\frac{2n^2-3n+2}{n(2n-1)r^{2n-1}}\right)\right]\label{eq:phoretic_21},\\
c'_1(\infty)=0,\,\,\quad&\label{eq:phoretic_22}\\
\totd{c'_1}{r}(1)=-\Da&\sum_{n=0}^\infty\frac{3k_nk_{n+1}}{(2n+1)(n+2)}\cdot\label{eq:phoretic_23}
\end{align}
Using Eq.~\eqref{eq:refsol} and solving for $c'_1(r)$, we finally obtain as an expansion at small $\Pe$ and $\Da$
\begin{equation}
U=\frac{k_1M}{3}+\Pe M^2H_1+\Da M H_2+o(\Da,\Pe)\label{eq:asymptU},
\end{equation}
where $H_1$ and $H_2$ are two constants that depend solely  on the details of the  surface activity distribution $k(\mu)$ as
\begin{align}
H_1&=-\frac{k_1k_0}{12}+\frac{k_1k_2}{90}+\sum_{n=2}^\infty\frac{(4n-1)k_nk_{n+1}}{(2n+1)(2n+2)(2n+3)(2n+4)}\label{eq:asymptUH1},\\
H_2&=-\sum_{n=0}^\infty\frac{k_nk_{n+1}}{(n+2)(2n+1)}\label{eq:asymptUH2}.
\end{align}
 Specifically, the sensitivity of the velocity to reactive ($\Pe$) and advective ($\Da$) effects is obtained as
\begin{align}
\left(\frac{1}{U}\pard{U}{\Pe}\right)_{(\textrm{Pe},\textrm{Da})=(0,0)}=\frac{3MH_1}{k_1},\qquad \left(\frac{1}{U}\pard{U}{\Da}\right)_{(\textrm{Pe},\textrm{Da})=(0,0)}=\frac{3H_2}{k_1}\label{eq:sensitivity}
\end{align}
and $k_1>0$ by convention for all Janus particles considered (reactive cap on the right).

\begin{figure}
\begin{center}
\begin{tabular}{cc}
\includegraphics[width=6.5cm]{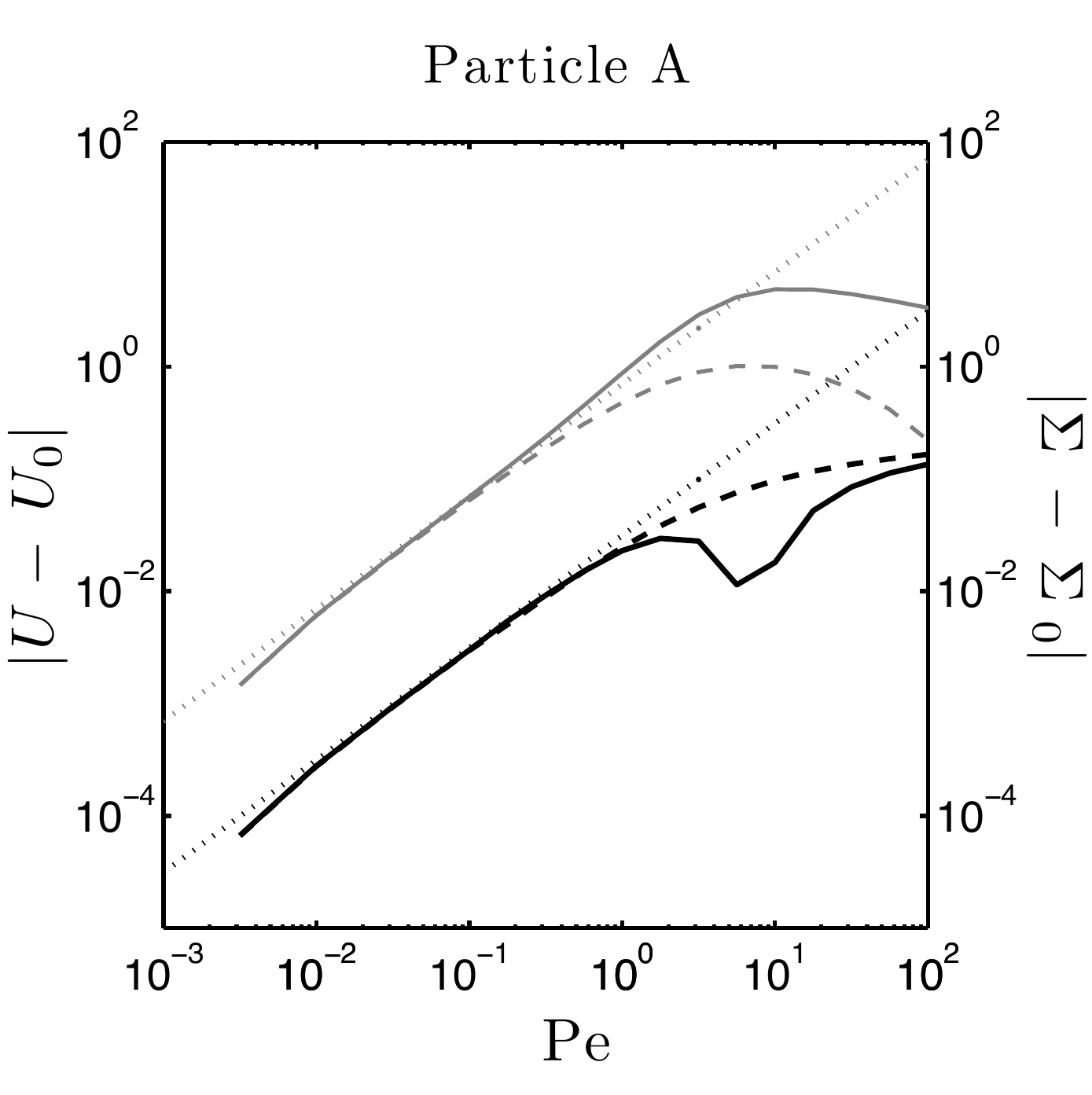} &
\includegraphics[width=6.5cm]{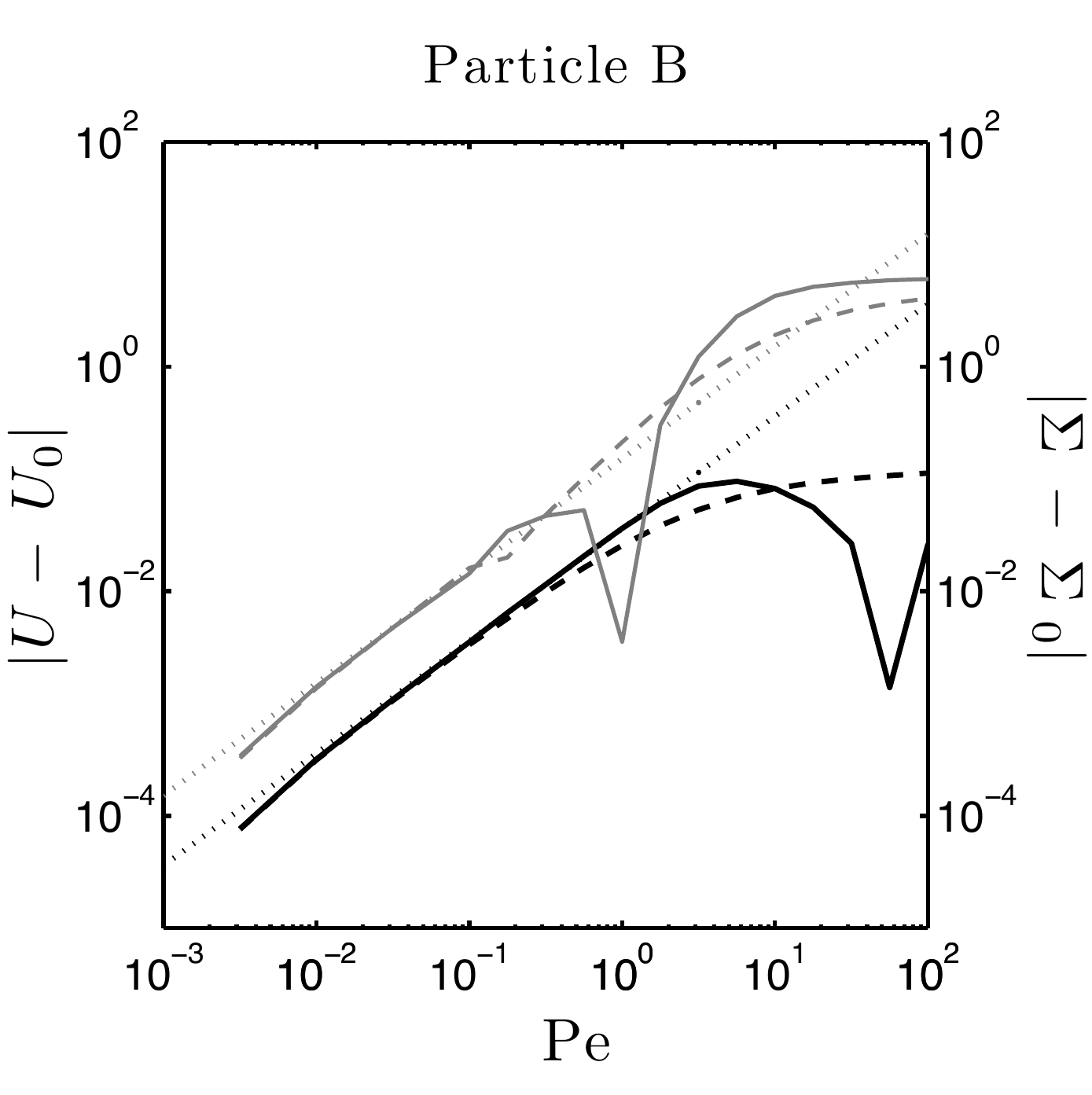} \\
\includegraphics[width=6.5cm]{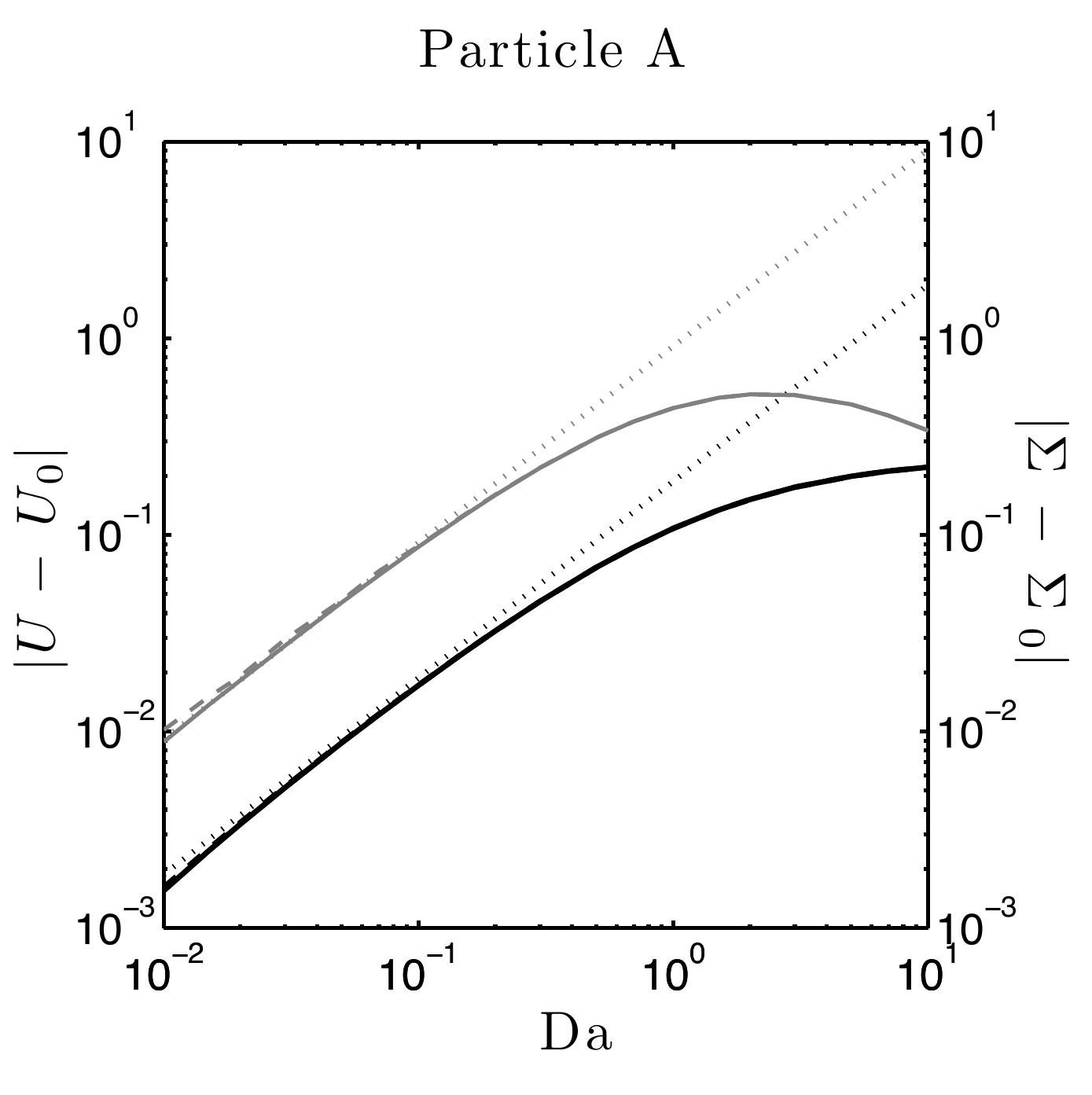} &
\includegraphics[width=6.5cm]{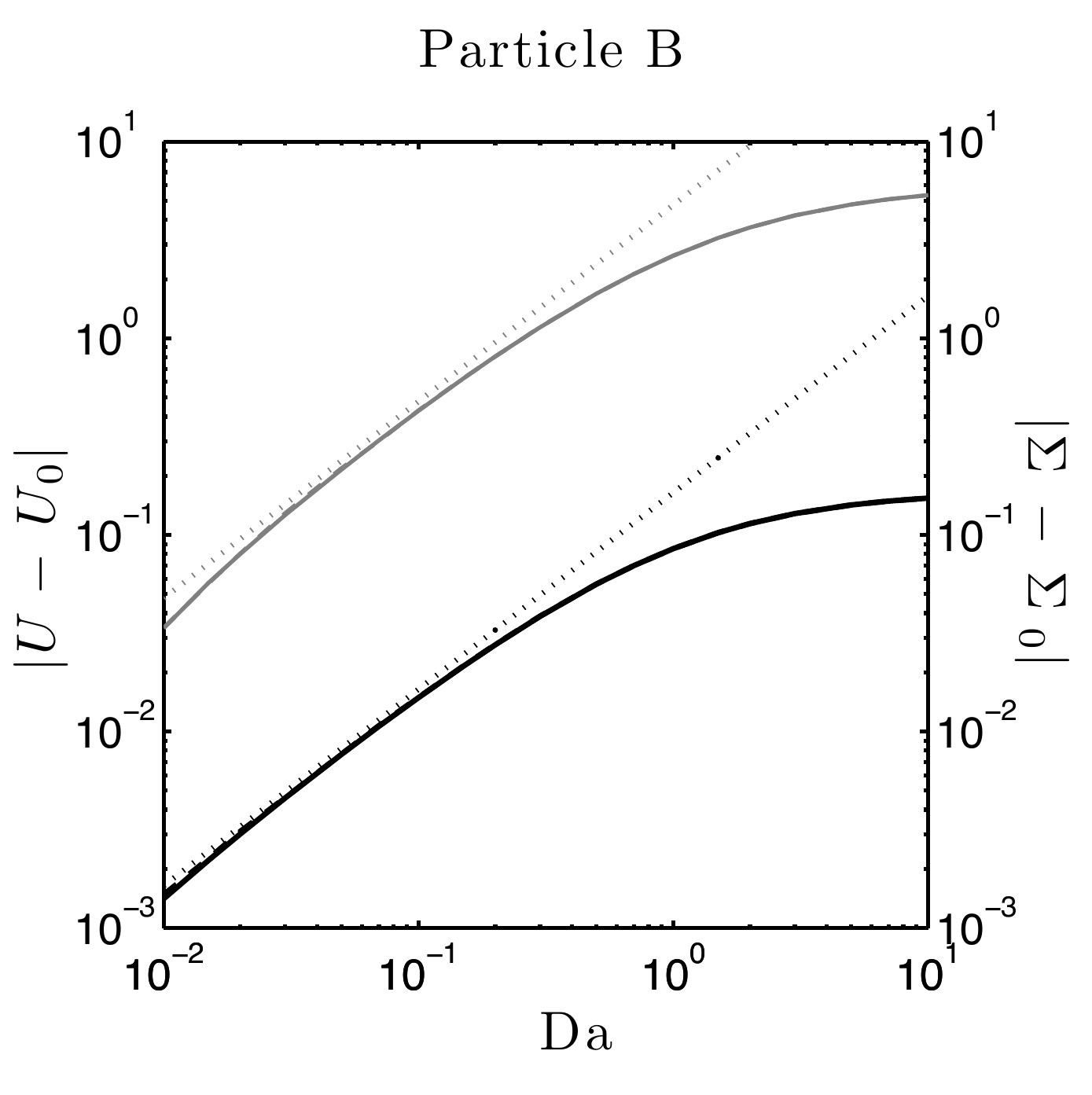}
\end{tabular}
\caption{\change{Comparison between the numerical results of Section~\ref{sec:results} with the asymptotic predictions for $\Pe,\Da\ll 1$ for particle A (left) and B (right): (Top) Evolution of $|U-U_0|$ (black) and $|\Sigma-\Sigma_0|$ (grey) with $\Pe$ for $\Da=0$, with $U_0$ and $\Sigma_0$ the phoretic velocity and stresslet intensity in the absence of advective or reactive effects; (Bottom) Evolution of the same quantities with $\Da$ for $\Pe=0$. For all figures, solid (resp.~dashed) lines correspond to particles with negative (resp.~positive) mobility. The dotted lines correspond to the predictions of 
Eqs.~\eqref{eq:asymptU} and \eqref{eq:asymptSig}.}  }\label{fig:comparaison_asymptotics}
\end{center}
\end{figure}

\begin{figure}
\begin{center}
\begin{tabular}{c}
\includegraphics[width=.6\textwidth]{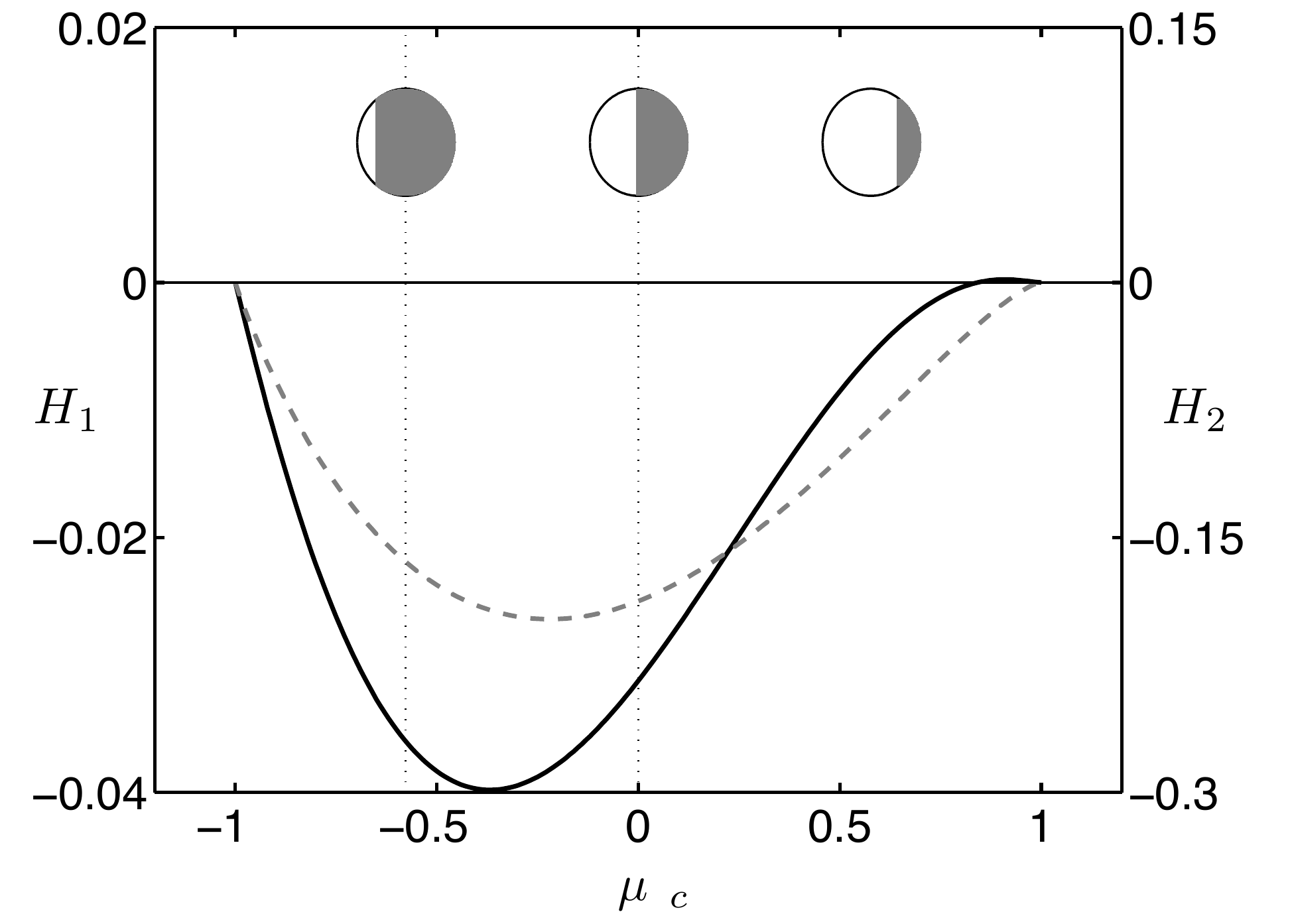} \\
\includegraphics[width=.6\textwidth]{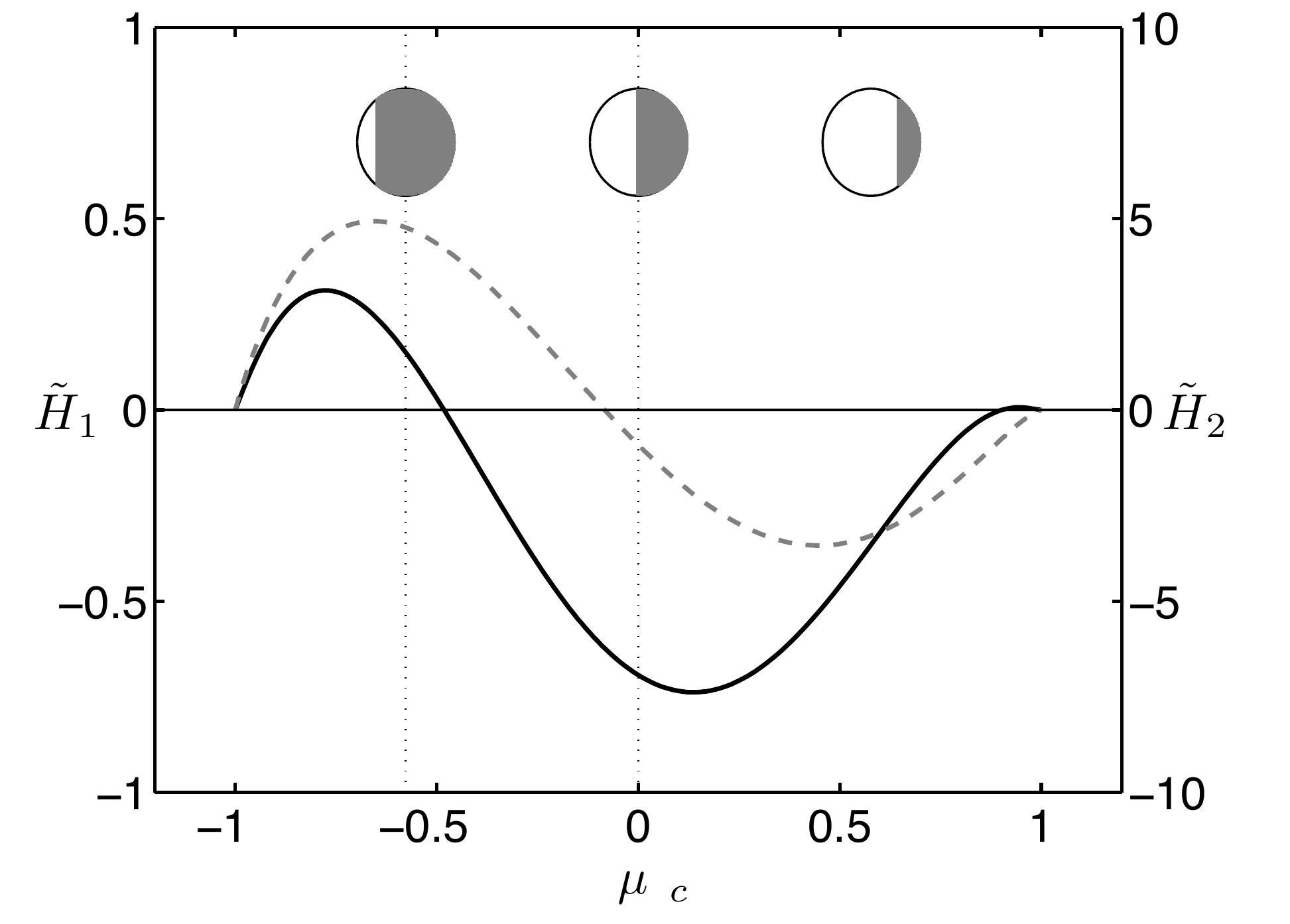} 
\end{tabular}
\caption{Top: Dependence of the sensitivities $H_1$ (solid) and $H_2$ (dashed) of the swimming velocity magnitude with $\Pe$ and $\Da$ for a Janus particle with  $k(\mu)=1_{[\mu_c, 1]}$ and positive mobility. Bottom: Evolution of the sensitivities $\tilde{H}_1$ (solid) and $\tilde{H}_2$ (dashed) of the stresslet intensity with $\Pe$ and $\Da$  for the same particles. Particles A and B are shown by dotted lines. }\label{fig:sensitivity}
\end{center}
\end{figure}

\change{This asymptotic  prediction shows an excellent agreement with the results obtained in Section~\ref{sec:results} for particles A and B (see Figure~\ref{fig:comparaison_asymptotics}).} For an arbitrary Janus particle, $k(\mu)=1_{[\mu_c,1]}$, the evolution of $H_1$ and $H_2$ with the size of the reactive cap (measured by $-1\leq\mu_c\leq 1$) is shown in Fig.~\ref{fig:sensitivity}. For all Janus particles, $H_1$ and $H_2$ are always negative, but their dependence is non-symmetric with respect to $\mu_c=0$ and two particles with reverse surface activity do not have the same sensitivity.

Since $H_2<0$ for all Janus particles, reactive effects always tend to reduce the velocity magnitude regardless of the sign of the mobility. This confirms our numerical observations  in the previous section for particles A and B, and can actually be extended easily  to any particle with a reactive ``stripe'' rather than a reactive ``cap'' (i.e. $k(\mu)=1_{[\mu_c,\mu_u]}$). Reactive effects systematically reduce the solute consumption rate near the active surfaces, effectively penalizing the chemical activity of that region by limiting the supply in fresh solute. 

In contrast, Eq.~\eqref{eq:sensitivity} shows that the sensitivity of the autophoretic  velocity to advective effects depends  on the sign of the mobility $M$. For all Janus particles, regardless of the size $\mu_c$ of the active region, a positive (resp.~negative) mobility leads  to  a reduction (resp.~increase) in their velocity magnitude from advective effects.  This extends to arbitrary $\mu_c$ our numerical results for particles A and B. When the particle swims toward its reactive pole ($M>0$), solute advection brings fluid of higher solute content closer to the reactive cap, reducing the contrast with the inert cap and the slip velocity magnitude. Instead, when the particle swims toward its inert pole ($M<0$), advective effects increase the solute content near the front cap (the inert one) and concentrates the depleted region near the reactive pole, increasing the tangential solute gradients and the slip velocity.

\subsection{Asymptotic analysis for the stresslet}

Following a similar approach for the stresslet and linearizing Eqs.~\eqref{eq:phoretic_11}--\eqref{eq:phoretic_13} for the $p=2$  mode, one obtains the asymptotic result 
\begin{equation}
\Sigma=4\pi M k_2+\Pe M^2\tilde{H}_1+\Da M \tilde{H}_2+o(\Da,\Pe),\label{eq:asymptSig}
\end{equation}
with
\begin{align}
\tilde{H}_1=10\pi&\left[-\frac{k_1^2+k_0k_2}{30}+\frac{k_1k_3}{112}+\sum_{n=2}^\infty\left(\frac{3n(n-2)k_n^2}{2(n+1)^2(2n-1)(2n+1)(2n+3)}\right.\right.\nonumber\\
&+\left.\left.\frac{3(3n-1)k_nk_{n+2}}{2(n+3)(2n+1)(2n+3)(2n+5)}\right)\right],\label{eq:asymptSigH1}\\
\tilde{H}_2=-4\pi&\sum_{n=0}^\infty\left[\frac{5n k_n^2}{(2n-1)(2n+1)(2n+3)}+\frac{15(n+2)^2k_nk_{n+2}}{(2n+1)(2n+3)(2n+5)(n+3)}\right]\cdot\label{eq:asymptSigH2}
\end{align}

\change{This asymptotic result is, once again, in excellent agreement with the numerical results of Section~\ref{sec:results} (Figure~\ref{fig:comparaison_asymptotics}). These results also emphasize the difference in the evolution of the stresslet for particles A and B. For particle A}, $k_2=0$ and $\Sigma_0=0$, and the asymptotic form in Eq.~\eqref{eq:asymptSig} is consistent with the stresslet at finite $\Pe$ (and $\Da=0$) being negative regardless of the sign of the mobility (Fig.~\ref{fig:Peeffect}). When $\Da\neq 0$, Eq.~\eqref{eq:asymptSig} also confirms that particles A of opposite mobility have stresslets of different sign (see Fig.~\ref{fig:Daeffect}).

When $\Sigma_0\neq 0$ (e.g. particle B in Section~\ref{sec:results}), the relative sensitivity of the stresslet is obtained as
\begin{equation}
\left(\frac{1}{\Sigma}\pard{\Sigma}{\Pe}\right)_{(\textrm{Pe},\textrm{Da})=(0,0)}=\frac{M\tilde{H}_1}{4\pi k_2},\qquad \left(\frac{1}{\Sigma}\pard{\Sigma}{\Da}\right)_{(\textrm{Pe},\textrm{Da})=(0,0)}=\frac{\tilde{H}_2}{4\pi k_2}.\label{eq:sensitivity2}
\end{equation}
For generic Janus particles that have a finite stresslet  in the reference configuration ($\Pe=\Da=0$), Eq.~\eqref{eq:sensitivity2} shows that increasing reactive effects ($\Da$) will have the same effect on particles of opposite mobility, while increasing advective effects will either lead to a maximum stresslet at intermediate $\Pe$ or a monotonic decrease of $\Sigma$ with $\Pe$, consistently with Figs.~\ref{fig:Peeffect} and \ref{fig:Daeffect} (note that the maximum in $\Sigma$ for particle B is of very small amplitude).

Finally, it should be noted that the results of the asymptotic analysis in Eqs.~\eqref{eq:asymptU}--\eqref{eq:asymptUH2} and \eqref{eq:asymptSig}--\eqref{eq:asymptSigH2} are not restricted to Janus activity distributions but  hold for any axisymmetric distribution of activity and could be used directly to investigate the sensitivity of a more general class of phoretic  particles.

\subsection{Optimal $\Pe$ for arbitrary Janus particles}

\begin{figure}
\begin{center}
\includegraphics[width=.7\textwidth]{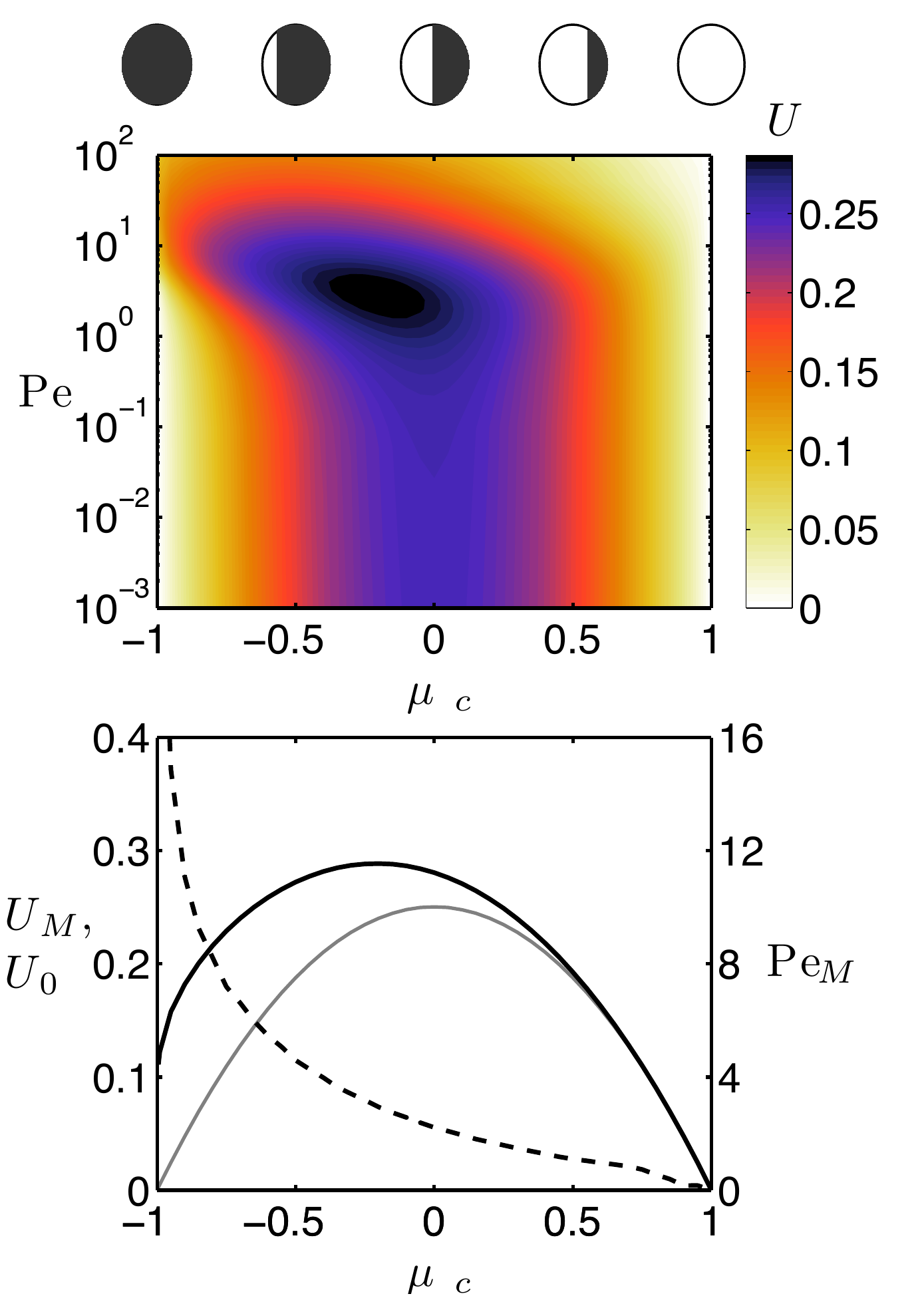}
\caption{(Colour online) Top: Dependence of the phoretic velocity magnitude on $\Pe$ and the relative size of the reactive cap $\mu_c$ (see illustration on  top); $\mu_c=-1$ corresponds to a fully reactive particle and $\mu_c=1$ corresponds to a fully inert particle. All reactive effects are neglected ($\Da=0$) and negative mobility $M=-1$ is considered, so the particles swim to the left. 
Bottom: Dependence on  $\mu_c$ of the optimal P\'eclet number, $\Pe_M$ (dashed), leading to the maximum velocity, $U_M$ (\change{black} solid), and of the self-propulsion velocity in the absence of advective effects, $U_0$  ($\Pe=0$, \change{grey} solid).}\label{fig:optimPejanus}
\end{center}
\end{figure}

The results in Fig.~\ref{fig:sensitivity} suggest that the sensitivity to advective effects is strongly dependent on the extent of the reactive cap for an arbitrary Janus particle of negative mobility $M=-1$, and is non-symmetric with respect to $\mu_c=0$. In particular, a maximum is reached for $\mu_c\approx -0.37$, with a sensitivity more than $25\%$ greater than the sensitivity to $\Pe$ of the symmetric particle A ($\mu_c=0$). The sensitivity to advective effects tends to vanish in the limit of isotropic inert or active particles ($\mu_c\rightarrow\pm 1$). Further, when comparing the sensitivity for two symmetric particles, i.e. those with inverse cap-size ratio or equivalently opposite $\mu_c$, the most reactive particle is always more sensitive to advective effects.  As $U\rightarrow 0$ when $\Pe\rightarrow\infty$ for all Janus particles, these results demonstrate the existence of an optimal $\Pe$ for $M<0$ leading to a maximum velocity.

This observation is confirmed using nonlinear numerical simulations and systematically   varying $\mu_c$ and $\Pe$.  The results are shown in Fig.~\ref{fig:optimPejanus} where we display the iso-values of the  phoretic velocity magnitude  (top) and plot the value of the maximum swimming velocity and corresponding optimal P\'eclet number as a function of the active cap size, $\mu_c$.  Note that $\Da=0$ is chosen here since we identified a systematic penalization of the swimming velocity by reactive effects.  The asymmetry between two symmetric particles $P_1$ and $P_2$ such that $k_{P_1}(\mu)=1-k_{P_2}(\mu)$ is apparent in 
Fig.~\ref{fig:optimPejanus}. Advective effects are observed to significantly enhance the swimming velocity of particles with $\mu_c<0$ (i.e. particles whose reactive cap is greater than the inert cap), while this effect is small for particles with smaller reactive caps ($\mu_c>0$). This observation is consistent with a monotonic decrease with $\mu_c$ of the optimal P\'eclet number, $\Pe_M$, at which   the maximum velocity is achieved. A mostly inert particle will experience a small peak velocity for low $\Pe$ while mostly active particles experience a large velocity increase for $\Pe\gtrsim 1$--$10$.

\begin{figure}
\begin{center}
\includegraphics[width=.6\textwidth]{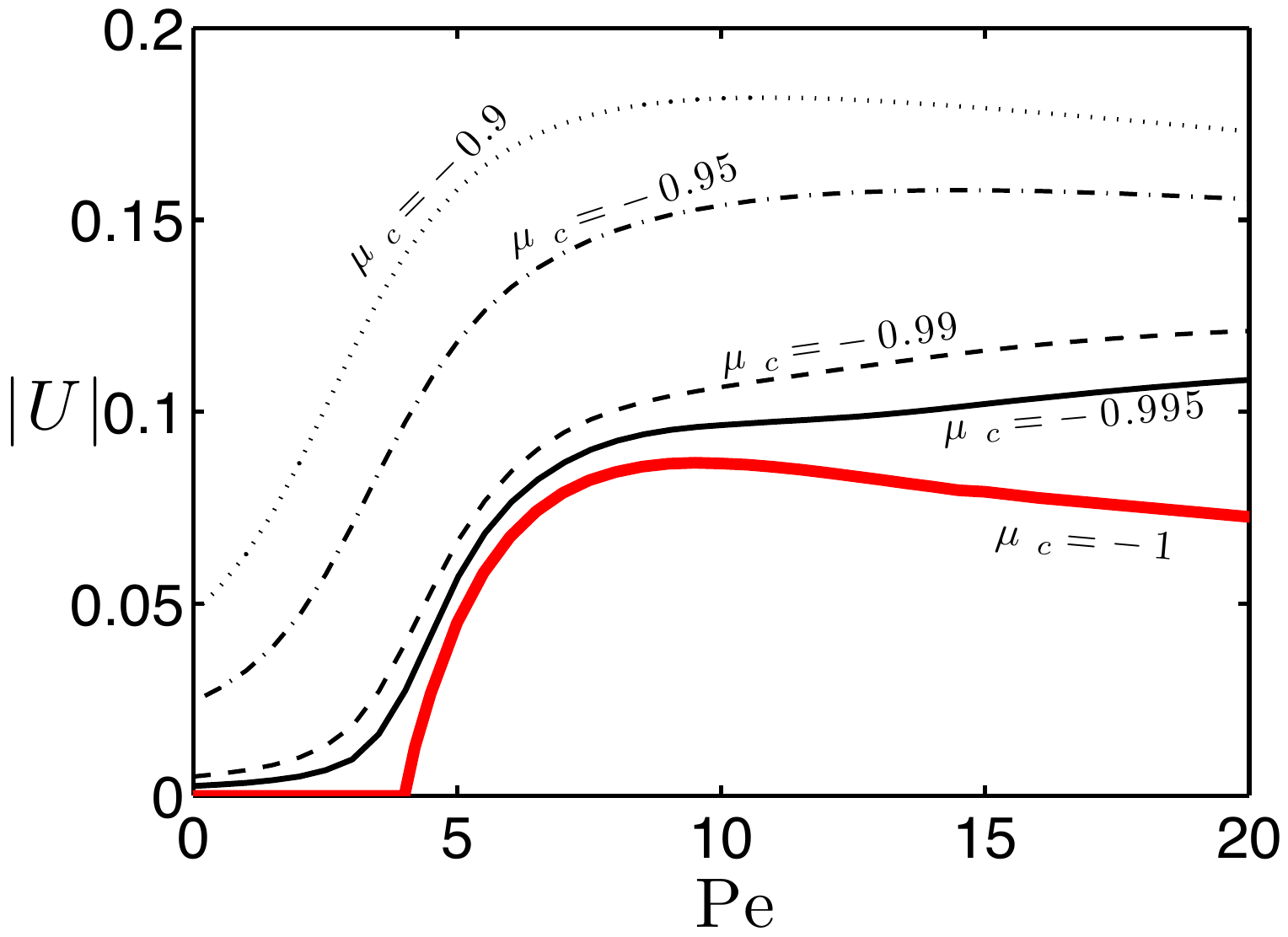}
\caption{(Colour online) Dependence  of the velocity magnitude, $|U|$,  with the P\'eclet number, $\Pe$, of Janus particles with 
 $\mu_c=-0.9$ (dotted line), 
$\mu_c=-0.95$ (dash-dotted),
$\mu_c=-0.99$ (dashed),
$\mu_c=-0.995$ (thin solid) with negative mobility. 
The  velocity obtained for a strictly isotropic particle $\mu_c=1$  and resulting from a symmetry-breaking instability \citep{michelin2013c} is shown in thick red line.}\label{fig:compare_isotropic}
\end{center}
\end{figure}

The limit $\mu_c\rightarrow -1$ is  particularly intriguing. This  corresponds to an almost fully-reactive particle except for a very small inert cap near the left pole. Such particles do not experience any significant self-propulsion at $\Pe=0$; however, Fig.~\ref{fig:optimPejanus} shows that such a particle may achieve a finite propulsion velocity for large $\Pe$. Ref.~\citep{michelin2013c} showed that the completely reactive particle ($\mu_c=-1$) may achieve self-propulsion at finite $\Pe$ despite spherical symmetry, through an instability and symmetry-breaking process arising  from the nonlinear coupling of the solute dynamics and phoretic slip velocity near the surface of the particle. The mechanism leading to self propulsion at high $\Pe$ has the same origin. More precisely, 
the dependence  of $|U|$ with $\Pe$  appears to converge asymptotically to that obtained for the isotropic reactive particle when $\mu_c\rightarrow -1$ (Figure~\ref{fig:compare_isotropic}). Infinitesimal velocities are obtained below the instability threshold $\Pe=4$ \citep[see derivation in][]{michelin2013c}  and are solely due to the symmetry-breaking introduced by the presence of a small inert cap on the left. Beyond $\Pe=4$, the instability resulting from the nonlinear coupling of the surface phoretic flows and solute advection-diffusion dominates and leads to finite swimming velocity.

\begin{figure}
\begin{center}
\includegraphics[width=.7\textwidth]{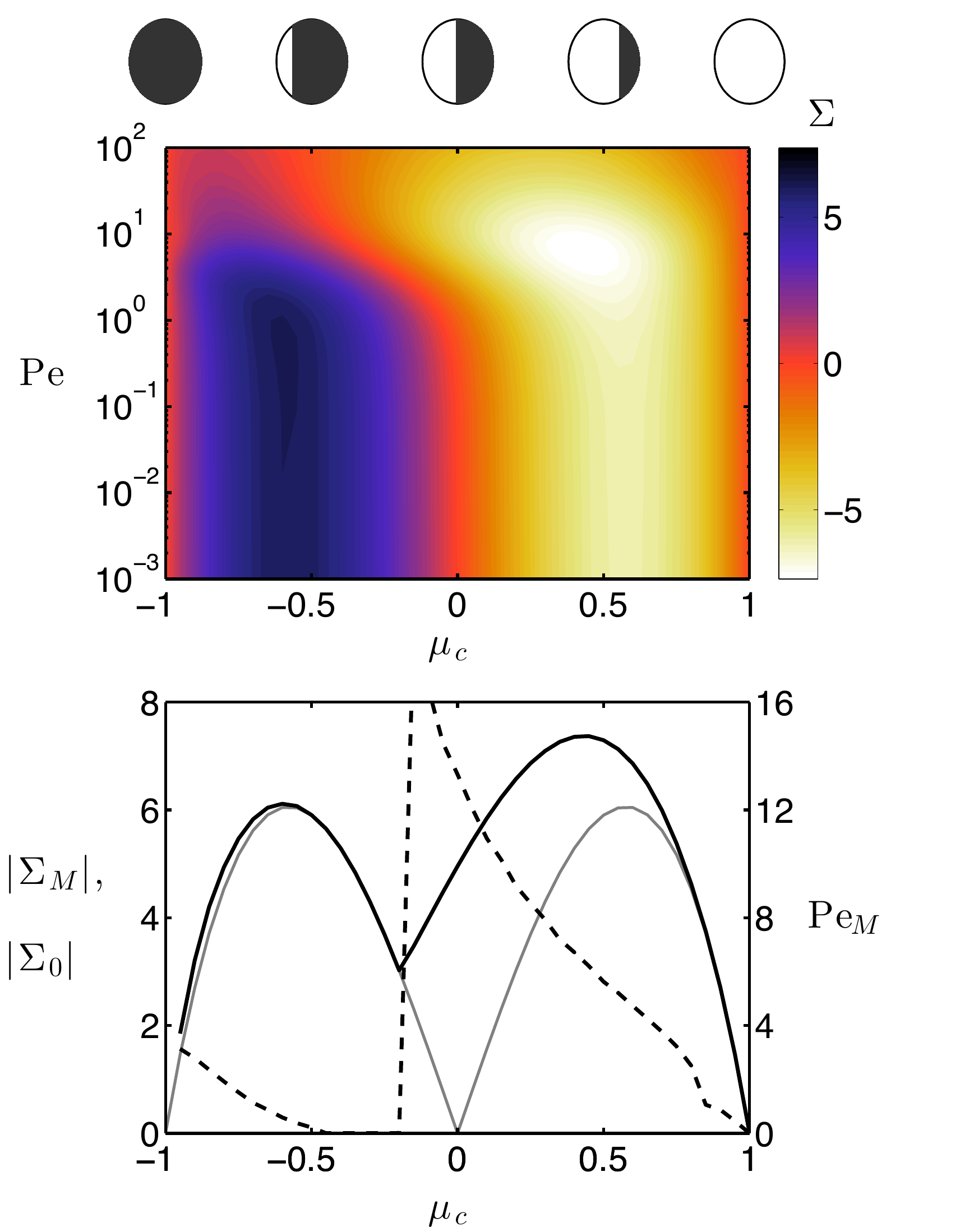}
\caption{(Colour online) Top: Dependence  of the stresslet intensity, $\Sigma$, with $\Pe$ and the relative size of the reactive cap, $\mu_c$. All reactive effects are neglected ($\Da=0$) and negative mobility $M=-1$ is considered (swimming occurs to the left). Bottom: Dependence on $\mu_c$ of the optimal P\'eclet number, $\Pe_M$ (dashed), leading to a maximum stresslet magnitude, $|\Sigma_M|$ (\change{black} solid), and of the stresslet magnitude in the absence of advective effects,  $|\Sigma_0|$ ($\Pe=0$, \change{grey} solid).}\label{fig:optimPejanus_stresslet}
\end{center}
\end{figure}

Finally, the dependence of the stresslet intensity, $\Sigma$, with $\Pe$ and $\mu_c$ also exhibits asymmetry, as shown in Fig.~\ref{fig:optimPejanus_stresslet}. At small $\Pe$, the sign of $\mu_c$  determines the sign of the stresslet. In the case of negative mobility  $M=-1$, mostly reactive particles behave as pullers ($\mu_c < 0$, $\Sigma>0$) while mostly inert particles behave as pushers 
($\mu_c >0$, $\Sigma<0$). The conclusions are reversed for $M=1$. At intermediate and large $\Pe$, this symmetry around $\mu_c=0$ no longer holds and  particles with a reactive cap slightly larger than a hemisphere  may experience a change in the sign of their stresslet intensity becoming pushers at large $\Pe$. For such particles, the maximum stresslet, resulting in the strongest inter-particle interaction, is obtained at large $\Pe$. Results obtained for $M=1$ (not shown here) exhibit the same dominance of the pusher characteristic: particles with $\mu_c>0$, which are pullers at $\Pe=0$, become pushers at larger $\Pe$.

\section{Conclusions}

Our work generalizes the classical continuum phoretic framework to account for finite advective and reactive effects. Our results highlight the influence of such effects on the self-propulsion of axisymmetric Janus particles, and are relevant to  phoretic particles of sufficiently large radius.  In particular, advection of the solute by the phoretic flows can lead to significant increases in the swimming velocity, which may reach a maximum at a finite, order one value of the  P\'eclet number. When the surface chemistry corresponds to a solute consumption at the particle  boundary, such a peak in the self-propulsion velocity is only observed for particles of negative mobility (corresponding to locally attractive solute-surface interactions), while particles of positive mobility experience a monotonic decrease of their propulsion velocity with $\Pe$. The impact of the geometrical active vs.~inert coverage of the particle surface was also identified. Particles that are predominantly reactive are more sensitive to advective effects and experience the largest increase in their swimming velocity at finite $\Pe$, while such effects are almost negligible for predominantly inert particles.

In contrast, reactive effects always penalize the self-propulsion of Janus phoretic  particles in all situations: when reaction acts too rapidly for the solute diffusion to refresh the solute content in the vicinity of the particle, the rate of consumption of the solute at the surface is reduced, effectively amounting to a reduction of surface activity that decreases the slip velocity and hence penalizes self-propulsion.

\change{Notably, the situation in which we predict locomotion to be enhanced  by advection effects (finite value of $\Pe$) is that of the system recently considered in a number of experimental investigations  \citep{Howse2007,Ebbens2011,ebbens2012}. In this setup, polymeric particles half coated with platinum are used to catalyze the autodegradation of hydrogen peroxyde, $\rm H_2O_2$, into dioxygen, $\rm O_2$. Self-diffusiophoresis due to the action of $\rm O_2$ gradients corresponds to net locomotion with the polymeric side of the particle first: $k < 0$ and $M> 0$. This situation is   equivalent to the case $k > 0$ and $M< 0$ computed above, and for which we predict locomotion to be enhanced by solute advection. Past experiments with this system have a P\'eclet number too small by about one order of magnitude for the effect to have been  observed yet, but it could play a role in the case of larger particles. It also certainly plays an important role in the case of motion driven by surfactant gradients where, due to much smaller molecular diffusivity,  P\'eclet numbers can easily reach $O(100)$ \citep{thutupalli2011}.}

In this paper, we chose to focus exclusively on propulsion through  self-diffusiophoresis of a particle, which catalyzes a simple one-step chemical reaction $S\rightarrow \emptyset$ on its surface. This framework can easily be extended to account for situations where both reagents and products of the reaction interact significantly with the surface or for multiple-steps reactions \citep[e.g. see][]{ebbens2012}. More generally, the results of our study are likely to remain valid for other phoretic mechanisms, provided the particle possesses two properties, namely \emph{mobility} and \emph{activity}. The former characterizes the ability of the particle to generate a slip velocity from an external field modified by advection and diffusion, while the latter corresponds to its ability to create local gradients of this field through chemical reaction or heat absorption/release \citep[e.g.][]{jiang2010,bickel2013}.  

Reactive and advective effects not only modify propulsion velocities but also significantly impact the flow field created by the particles and is  therefore expected to modify the type and intensity of its hydrodynamic interactions with neighboring particles. The collective dynamics of such colloid particles was shown recently to exhibit different complex behaviours \citep{theurkauff2012,palacci2013}. A complete fundamental understanding of the mechanisms leading to such aggregations and collective organizations remains to be obtained. The results presented in this work however suggest that for larger particles, advective and, to a smaller extent, reactive effects may be significant.

Finally, the results obtained here rely on the assumption of a thin interaction layer, considered in most existing literature. We carefully identified the conditions under which this assumption and resulting framework are valid, namely when advection within the interaction layer can be neglected. The generalization of these results to the limit of very large $\Pe$ (i.e. when $\varepsilon\Pe=O(1)$ or above) remains an open question. Depending on the nature of the solute-particle interaction, advection within the interaction layer may enhance or reduce the local solute gradients, respectively increasing or penalizing the self-propulsion velocity. \change{Another interesting question to consider would be the effect of interactions between solute molecules, effectively introducing variations of the interaction potential with the solute concentration.}

\appendix
\section{Generalization to non-uniform mobility}
\label{sec:nonunif_mob}
The framework and results presented in the main part of the paper   can easily be extended to the case of a particle with non-uniform mobility which may be more relevant to experimental conditions. 
Indeed,  the chemical treatment of the surface of the particle is likely to affect {both} surface activity and  mobility. From a theoretical point of view, the phoretic problem formulation in Eqs.~\eqref{eq:phoretic_01}--\eqref{eq:phoretic_05} remains valid for non-uniform $M(\mu)$. Its formulation in the squirmer framework is only marginally modified and while Eqs.~\eqref{eq:phoretic_11}--\eqref{eq:phoretic_13} remain unchanged,  Eq.~\eqref{eq:phoretic_14} must be replaced by
\begin{equation}
\alpha_n=-\sum_{m,p=0}^\infty\frac{n(n+1)}{(2n+1)(2p+1)}M_pc_m(1)B_{mnp},
\end{equation}
with $B_{mnp}$ defined in Eq.~\eqref{eq:defB} and $M(\mu)=\sum M_pL_p(\mu)$. 
We briefly show here that the results presented in the paper do remain valid when non-uniform mobility is considered,  focusing on the configuration $M(\mu)=\pm k(\mu)$ (the chemical patterning of the particle surface impacts both properties at the same time). 

\begin{figure}
\begin{center}
\includegraphics[width=.65\textwidth]{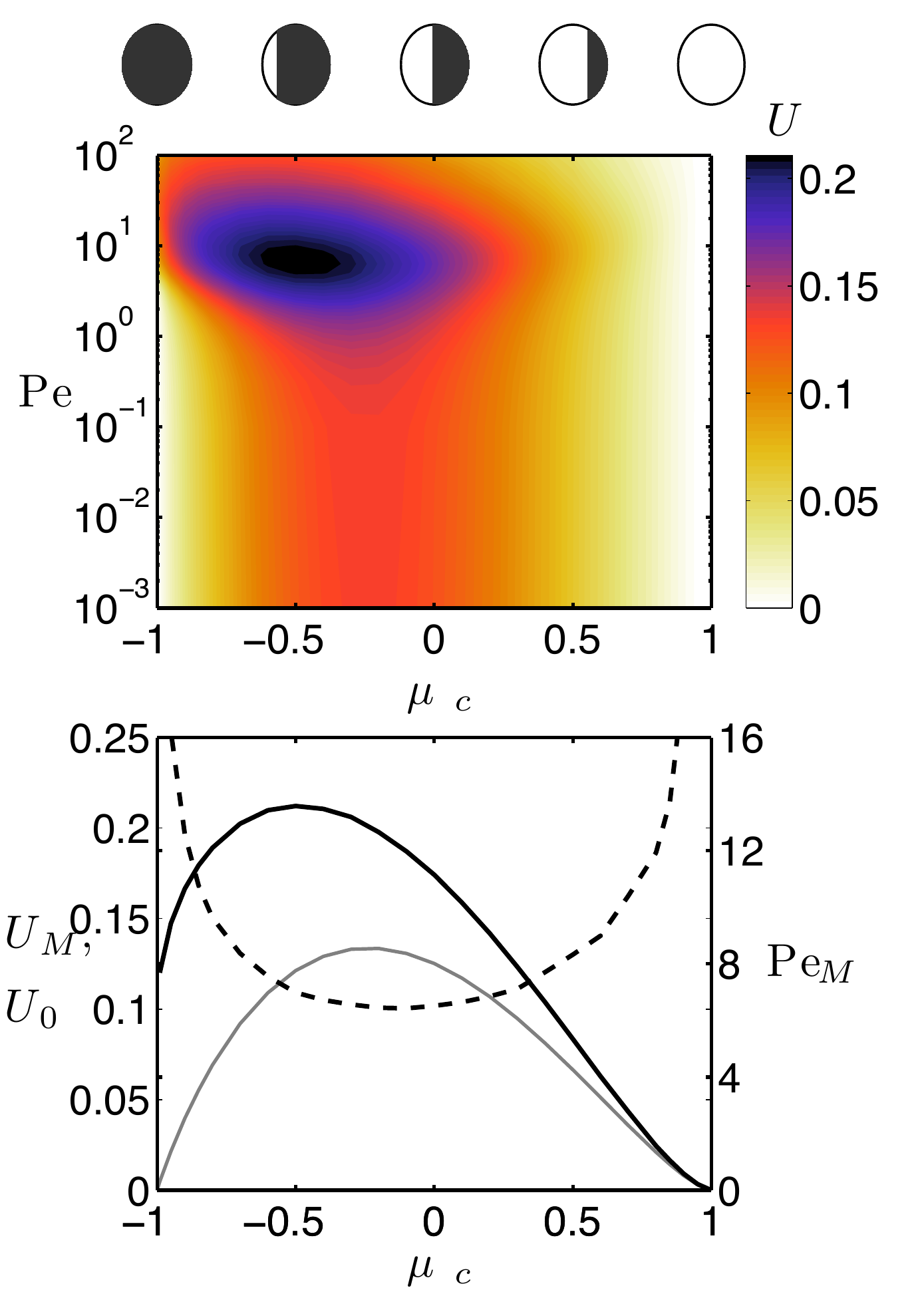}
\caption{Same results as Fig.~\ref{fig:optimPejanus} but for a particle with non-uniform mobility $M(\mu)=-k(\mu)$. 
Top: Dependence of the phoretic velocity magnitude with $\Pe$ and the relative size of the reactive cap, $\mu_c$. Reactive effects are neglected ($\Da=0$) and negative mobility $M(\mu)=-k(\mu)$ is considered (swimming is thus to the left). 
Bottom: Dependence on $\mu_c$ of the optimal P\'eclet number,
 $\Pe_M$ (dashed), leading to the maximum velocity, 
 $U_M$ (\change{black} solid), and of the self-propulsion velocity in the absence of advective effects,  $U_0$ ($\Pe=0$, \change{grey} solid).}\label{fig:optimalPe_varymob}
\end{center}
\end{figure}

Figure~\ref{fig:optimalPe_varymob} displays  the dependence of the swimming velocity of the particle with $\Pe$ and the surface chemical coverage and generalize thus the results of Fig.~\ref{fig:optimPejanus} to the case of non-uniform mobility. The swimming velocity levels are generally obviously reduced since a smaller fraction of the particle contributes to the slip velocity, but the main conclusions of the paper are  unchanged. Specifically, in the case of negative mobility, there exists an optimal $\Pe(\mu_c)$ maximizing the swimming velocity. Further, this advective effect is  more pronounced for particles whose reactive cap is greater than a hemisphere ($\mu_c<0$). The analysis obtained for $\mu_c\rightarrow -1$ is also strictly equivalent to the uniform mobility case: in that limit, the existence of a small cap with zero mobility near the left pole does not modify the swimming velocity significantly since it corresponds to a small surface and the slip velocity in that region is almost orthogonal to the swimming direction.

These results only differ from the ones in  the main paper on two minor points. First the swimming velocity at $\Pe=0$ is no longer symmetric with respect to $\mu_c=0$ and the optimal $\Pe$ for $\mu_c\rightarrow 1$ is large instead of converging to 0. The former result is a consequence of the greater surface contributing to the swimming velocity when $\mu_c<0$ in comparison with $\mu_c>0$. The latter indicates that the optimal velocity is reached for large $\Pe$. However, as for the configuration with uniform mobility, this optimal velocity only marginally differs from the reference velocity $U_0$ and such Janus particles are still only weakly sensitive to advective effects. We  finally note  that for non-uniform mobility, advective effects can lead to stronger increases in the swimming velocity.


\end{document}